\documentclass[aps,onecolumn,showpacs]{revtex4}
\usepackage{graphicx}
\usepackage{bm}
\usepackage{multirow}
\usepackage{array}
\usepackage{graphicx}
\usepackage{amsmath}
\usepackage{txfonts}
\usepackage{mathrsfs}
\usepackage{subfigure}
\usepackage{amssymb}
\usepackage{caption}
\usepackage{float}
\usepackage{enumerate}
\usepackage{color}
\definecolor{mygreen}{cmyk}{1,0.02,0.,0.42}        % dark green  
\definecolor{myblue}{cmyk}{0.38,0.18,0.,0.46}

\usepackage{slashed}
\usepackage{lipsum}
\usepackage{hyperref}
\usepackage{url}
\hypersetup{hypertex=true,
	colorlinks=true,
	linkcolor=mygreen,
	anchorcolor=red,
	citecolor=blue}
\begin{document}
	\setlength{\textfloatsep}{5pt}
	\title{Gauge independence of pion masses in a magnetic field within the Nambu--Jona-Lasinio model}
	\author{Jianing Li$^1$\footnote{ljn18@mails.tsinghua.edu.cn}} \author{Gaoqing Cao$^2$\footnote{caogaoqing@mail.sysu.edu.cn}} \author{Lianyi He$^1$\footnote{lianyi@mail.tsinghua.edu.cn}}
	\address{$^1$Department of Physics, Tsinghua University, Beijing 100084, China\\$^2$School of Physics and Astronomy, Sun Yat-sen University, Zhuhai 519088, China}
	\date{\today}
	
\begin{abstract}
We investigate the properties of neutral and charged pions in a constant background magnetic field mainly at zero temperature within the Nambu--Jona-Lasinio model. In the previous calculations, the Ritus method, involving Schwinger phases in a fixed gauge, was employed within the momentum-space random phase approximation (RPA)~[Phys. Lett. B $\textbf{782}$, 155-161 (2018)].  However, gauge invariance of the charged pion masses has not yet been examined. In this work,  by adopting the linear response theory based on the imaginary-time path integral formalism, we derive the correlation functions for pions in the coordinate space, where the corresponding Schwinger phases show up automatically. At sufficiently large imaginary time $\tau$, the meson correlation function approaches an exponential form $\sim\exp(-E_{\rm G}\tau)$, where $E_{\rm G}$ is the ground-state energy of the one-meson state and hence determined as the meson mass. Furthermore, we show that the mass of the charged pions is gauge independent, i.e., independent of the choice of the vector potential for the magnetic field. Actually, we also find that the momentum-space RPA is equivalent to the imaginary-time method used here.
\end{abstract}
	
\maketitle
	
%%%%%%%%%%%%%%%%%%%%%%%%%%%%%%%%%%%%%%%%%%%%%%%%%%%%%%%%%%%%%%%%%%%%%%%%%%%%%%%%%%%%%%%%%%%%%%%%%%%%%%%%%%%%%%%%%
\section{Introduction}
%%%%%%%%%%%%%%%%%%%%%%%%%%%%%%%%%%%%%%%%%%%%%%%%%%%%%%%%%%%%%%%%%%%%%%%%%%%%%%%%%%%%%%%%%%%%%%%%%%%%%%%%%%%%%%%%%
The properties of quantum chromodynamics (QCD) matter in a strong magnetic field have recently attracted numerous attentions in high energy nuclear physics~\cite{Kharzeev:2013jha,Andersen:2014xxa,Miransky:2015ava,Huang:2015oca}. The significance of this topic is mainly relevant to strong magnetic fields found in various real systems related to QCD: the surface of magnetars~\cite{Duncan:1992hi,Baym:2019iky}, the inner core of pulsars~\cite{Camilo:2000mp,Blaschke:1999fy}, and the fireballs produced in peripheral heavy ion collisions~\cite{Skokov:2009qp,Voronyuk:2011jd,Deng:2012pc}. The external magnetic field serves as an extra dimension and enriches the QCD phase diagram~\cite{Andersen:2014xxa,Miransky:2015ava,Bali:2011qj,Cao:2021rwx}. Meanwhile,  magnetic-field induced effects have been proposed theoretically and can be probed in recent or future experiments, such as chiral magnetic effect~\cite{Fukushima:2008xe,Kharzeev:2013ffa}, neutral pion condensation~\cite{Cao:2015cka}, and disputable superconductivity in magnetized vacuum~\cite{Chernodub:2010qx,Hidaka:2012mz,Cao:2019res}.
	
The chiral symmetry breaking or restoration is one of the most important aspects of QCD under extreme circumstances. At zero temperature, it was proposed that an external magnetic field enhances the chiral condensate, known as the magnetic catalysis (MC) effect~\cite{Bali:2012zg,Shovkovy:2012zn}. However, around the critical temperature, lattice QCD calculations found that the external magnetic field reduces the chiral condensate, which is now called the inverse magnetic catalysis (IMC) effect~\cite{Bruckmann:2013ufa,Bruckmann:2013oba,Bandyopadhyay:2020zte}. On the other hand, the meson properties in a magnetized QCD system have been studied extensively. In the massless limit, the SU$(2)$ chiral symmetry of two-flavor  QCD  is explicitly broken down to a U$(1)$ subgroup by the external magnetic field, and thus only the neutral pion is the Goldstone boson associated with the spontaneous breaking of the residual chiral symmetry. The properties of both neutral and charged pions in a constant magnetic field at zero and finite temperature have been studied by utilizing lattice QCD (LQCD) simulations~\cite{Hidaka:2012mz,Luschevskaya:2015bea,Luschevskaya:2015cko,Bali:2015vua,Bali:2017ian,Ding:2020jui,Ding:2020hxw}, chiral perturbation theory~\cite{Kamikado:2013pya,GomezDumm:2017jij}, chiral effective models including linear sigma model~\cite{Ayala:2018zat,Das:2019ehv,Ayala:2020dxs} and Nambu--Jona-Lasinio (NJL) model~\cite{Xu:2020yag,Avancini:2016fgq,Wang:2017vtn,Mao:2017wmq,Liu:2018zag,Mao:2018dqe,Coppola:2018vkw,Coppola:2019uyr,Avancini:2018svs,Sheng:2020hge,Ghosh:2020qvg} as well as its Polyakov-loop extension~\cite{Dumm:2020muy}, and other effective models~\cite{Andreichikov:2016ayj,Aguirre:2017dht}. For the neutral pion, there is no ambiguity on the definition of its pole mass as the Schwinger phase~\cite{Schwinger:1951nm} vanishes. It was found that the neutral pion mass is reduced by a weak magnetic field, whereas its tendency in stronger magnetic field is still uncertain~\cite{Hidaka:2012mz,Luschevskaya:2015bea,Luschevskaya:2015cko,Bali:2015vua,Bali:2017ian,Ding:2020jui,Ding:2020hxw,Xu:2020yag,Avancini:2016fgq,Wang:2017vtn,Mao:2017wmq,Liu:2018zag,Mao:2018dqe,Coppola:2018vkw,Coppola:2019uyr,Avancini:2018svs,Sheng:2020hge,Ghosh:2020qvg}. For charged pions, the non-vanishing Schwinger phase makes it hard to perform a complete momentum-space calculation. Nevertheless, in a recent work~\cite{Coppola:2018vkw}, the Schwinger phases of charged pions have been taken into account via the Ritus method and a monotonical increase of the charged pion mass with the magnetic field strenth was found. Recent lattice QCD results~\cite{Ding:2020jui,Ding:2020hxw}, however, showed that the charged pion mass starts to decrease at sufficiently large magnetic field.
	
So far, most of the investigations of the charged pion properties were performed in a fixed gauge, i.e., the Landau gauge or the symmetric gauge. Before we seek for the way out to understand the lattice results for the charged pions, the gauge independence of the pion masses needs to be examined preferentially. In this work, we investigate the pion masses in a constant magnetic field and their gauge independence within the NJL model. In the framework of the NJL model, mesons are regarded as collective excitations of quark-antiquark pairs and can be constructed by using the random phase approximation, which guarantees the Goldstone theorem~\cite{Nambu:1961tp,Nambu:1961fr,Klevansky:1992qe}. In the absence of external magnetic field, the RPA is easy to be performed in the momentum space. However, in an external magnetic field, momenta are no longer good quantum numbers for charged mesons. In this case, the presence of a nonvanishing Schwinger phase renders the RPA impossible to be carried out in the momentum space.  Modified schemes to apply the RPA include discarding the Schwinger phase~\cite{Liu:2018zag,Cao:2019res}, local expansion of the Schwinger phase~\cite{Wang:2017vtn}, and the Ritus eigenfunction method~\cite{Coppola:2018vkw}. 

In this work, we perform the RPA directly in the coordinate space, inspired by recent lattice QCD calculations~\cite{Luschevskaya:2015cko,Ding:2020jui,Ding:2020hxw}. The two-point correlation function of a meson in coordinate space can be derived by using the path integral formulation of the linear response theory~\cite{He:2016cjd,Mu:2018wpf}. Applying imaginary-time path integral and integrating over the spatial coordinates, the correlation function approaches an exponential form $\sim\exp(-E_{\rm G}\tau)$ at sufficiently large $\tau$.  The ground-state energy $E_{\rm G}$ of the one-meson state thus determines the meson mass and is shown to be gauge independent within the NJL model, regardless of the choice of regularization scheme. We show numerically and analytically that the previously used momentum-space RPA, which simply discarded the Schwinger phase for charged pions and determined the pion masses as the poles at zero momentum, is actually equivalent to the imaginary-time method used in this work. Note that a recent work in the linear sigma model showed that the magnetic-field-induced vertex modification is also gauge invariant even though additional Schwinger phase dependence of the quark-meson coupling was taken into account~\cite{Ayala:2020muk}.
	
The paper is organized as follows. In Sec.~\ref{sec2}, we show that the meson mass can be defined via the meson correlation function at large imaginary time and establish a theoretical framework to calculate the meson correlation functions in a constant magnetic field in the two-flavor NJL model. In Sec.~\ref{sec3}, we calculate the correlation function of the neutral pion, from which the neutral pion mass is extracted. In Sec.~\ref{sec4}, the correlation function and mass of the charged pions are studied. A general proof of the gauge independence of charged meson masses and the equivalence between the momentum-space RPA  and the imaginary-time method is also presented. We summarize in Sec.~\ref{sec5}.

%%%%%%%%%%%%%%%%%%%%%%%%%%%%%%%%%%%%%%%%%%%%%%%%%%%%%%%%%%%%%%%%%%%%%%%%%%%%%%%%%%%%%%%%%%%%%%%%%%%%%%%%%%%%%%%%%
\section{Pion correlation functions}\label{sec2}
%%%%%%%%%%%%%%%%%%%%%%%%%%%%%%%%%%%%%%%%%%%%%%%%%%%%%%%%%%%%%%%%%%%%%%%%%%%%%%%%%%%%%%%%%%%%%%%%%%%%%%%%%%%%%%%%%
We start from the imaginary-time correlation function for a mesonic state in a magnetized QCD vacuum,
\begin{equation} 
{\cal D}_{\rm M}(\tau,\mathbf{r};\tau^\prime,{\bf r}^\prime)\equiv\langle{\rm vac} |{\rm T}_\tau \hat{\phi}(\tau,{\bf r})\hat{\phi}^\dagger(\tau^\prime,{\bf r}^\prime)|{\rm vac}\rangle,
\end{equation}
with $|{\rm vac}\rangle$ being the vacuum state of the system, $\tau$ being the imaginary time, and ${\rm T}_\tau$ denoting the imaginary-time order. The composite field operator for the mesonic state, 
$\hat{\phi}(\tau,{\bf r})$, is constructed by using the quark field operator $\hat{\psi}$,
\begin{equation} 
\hat{\phi}(\tau,{\bf r})=\hat{\bar{\psi}}(\tau,{\bf r})\Gamma_{\rm M}\hat{\psi}(\tau,{\bf r}),
\end{equation}
where the matrix $\Gamma_{\rm M}$ characterizing the mesonic state can be decomposed as $\Gamma_{\rm M}=\Gamma_{\rm D}\otimes\Gamma_{\rm F}$, with $\Gamma_{\rm D}$ and $\Gamma_{\rm F}$ being matrices in the
spin and flavor spaces, respectively. In this work, we are interested in the pions and hence $\Gamma_{\rm D}=i\gamma_5$.

Because of the translational invariance in the temporal dimension,  we set $\tau^\prime=0$ without loss of generality. Since we are interested in the limit $\tau\rightarrow\infty$, we focus on the case $\tau>0$. In this case, we have
\begin{eqnarray} 
{\cal D}_{\rm M}(\tau,{\bf r},{\bf r}^\prime)\equiv{\cal D}_{\rm M}(\tau,{\bf r};0,{\bf r}^\prime)=\langle{\rm vac} | e^{\tau\hat{H}}\hat{\phi}_{\rm S}^{\phantom{\dag}}({\bf r})e^{-\tau\hat{H}}\hat{\phi}_{\rm S}^\dagger({\bf r}^\prime)|{\rm vac}\rangle,
\end{eqnarray}
where $\hat{H}$ is the Hamiltonian of the system, $\hat{\phi}_{\rm S}^{\phantom{\dag}}({\bf r})\equiv \hat{\phi}(0,{\bf r})$ is the composite field operator in the Schr\"oedinger picture, and we have used the time evolution
$\hat{\phi}(\tau,{\bf r})=e^{\tau\hat{H}}\hat{\phi}(0,{\bf r})e^{-\tau\hat{H}}$. The vacuum state is the ground state of the Hamiltonian $\hat{H}$, with the eigen-energy being the vacuum energy $E_{\rm vac}$, i.e., 
$e^{\tau\hat{H}}|{\rm vac}\rangle=e^{\tau E_{\rm vac}}|{\rm vac}\rangle$. Since the following derivations are only related to the one-meson states, we can set $E_{\rm vac}=0$ without loss of generality.

Now the correlation function becomes
\begin{eqnarray} 
{\cal D}_{\rm M}(\tau,{\bf r},{\bf r}^\prime)=\langle{\rm vac} | \hat{\phi}_{\rm S}^{\phantom{\dag}}({\bf r})e^{-\tau\hat{H}}\hat{\phi}_{\rm S}^\dagger({\bf r}^\prime)|{\rm vac}\rangle.
\end{eqnarray}
The composite field operator $\hat{\phi}_{\rm S}^\dagger({\bf r})$ acting on the vacuum state is only related to the one-meson states. Denoting the one-meson states as $|{\rm M}_l\rangle$, with $l$ being the collection of quantum numbers, we can write
\begin{eqnarray} 
\hat{\phi}_{\rm S}^\dagger({\bf r})|{\rm vac}\rangle=\sum_l f_l^{\phantom{\dag}}({\bf r})|{\rm M}_l\rangle,\ \ \ \ \ \ 
\langle{\rm vac} | \hat{\phi}_{\rm S}^{\phantom{\dag}}({\bf r})=\sum_{l}f_l^*({\bf r})\langle {\rm M}_l|.
\end{eqnarray}
The expansion coefficients $f_l({\bf r})$ are not important for the determination of the meson mass. The one-meson states $|{\rm M}_l\rangle$ are also the eigen-states of the Hamiltonian, with eigen-energies $E_l$. Therefore, the meson correlation function can be expressed as
\begin{eqnarray} \label{CORR-M}
{\cal D}_{\rm M}(\tau,{\bf r},{\bf r}^\prime)=\sum_{l,l^\prime}f_l^*({\bf r})f_{l^\prime}^{\phantom{\dag}}({\bf r}^\prime)\langle {\rm M}_l|{\rm M}_{l^\prime}\rangle e^{-\tau E_{l^\prime}}
=\sum_{l}f_l^*({\bf r})f_{l}^{\phantom{\dag}}({\bf r}^\prime) e^{-\tau E_l},
\end{eqnarray}
where we have used the fact $\langle {\rm M}_l|{\rm M}_{l^\prime}\rangle=\delta_{ll^\prime}$.

The meson mass $m_{\rm M}$ is defined as the lowest eigen-energy of the one-meson state, i.e., 
\begin{eqnarray} 
m_{\rm M}=E_{\rm G}\equiv \min_l \{E_l\}.
\end{eqnarray}
Eq. (\ref{CORR-M}) shows that at large positive $\tau$ ($\tau\rightarrow\infty$), only the term with the lowest energy $E_{\rm G}$ survives, that is
\begin{eqnarray}\label{CORR-M0}
{\cal D}_{\rm M}(\tau,{\bf r},{\bf r}^\prime)\rightarrow f_{\rm G}^*({\bf r})f_{\rm G}^{\phantom{\dag}}({\bf r}^\prime) e^{-\tau E_{\rm G}}
\end{eqnarray}
Therefore, we can utilize the exponential form $\exp(-E_{\rm G}\tau)$ at large positive $\tau$ to extract the meson mass $m_{\rm M}$.  For a system with translational invariance, the correlation function is only a function of the 
relative coordinate ${\bf r}-{\bf r}^\prime$. We can set ${\bf r}^\prime=0$ and define 
\begin{eqnarray} \label{CORR-M1}
{\cal P}_{\rm M}(\tau)\equiv \int d^3{\bf r}{\cal D}_{\rm M}(\tau,{\bf r},{\bf 0}).
\end{eqnarray}
Since the exponential form $\exp(-E_{\rm G}\tau)$ at large positive $\tau$ is not related to the translational invariance, for a system without translational invariance, we can still set ${\bf r}^\prime=0$ and define 
${\cal P}_{\rm M}(\tau)$
in the same way. In any case, we can extract the meson mass $m_{\rm M}=E_{\rm G}$ from the large-$\tau$ behavior ${\cal P}_{\rm M}(\tau)\sim\exp(-E_{\rm G}\tau)$.

On the other hand,  the correlation function can also be defined and computed from the imaginary-time path integral formalism. To this end, we define the partition function in the presence of external sources,
\begin{equation}\label{partitionF}
{\cal Z}[J]=\int {\cal D}[\bar{\psi},\psi,\cdots]\exp{\left[\int {\rm d}^4X\left({\cal L}+J\phi+J^\dagger\phi^\dagger\right)\right]},
\end{equation}
where $\phi=\bar{\psi}\Gamma_{\rm M}\psi$ and $\int {\rm d}^4X\equiv \int {\rm d}\tau \int{\rm d}^3{\bf r}$. We will also use the notation $X=(\tau,{\bf r})$ in the following. While the explicit form of the partition function replies on the effective model we adopt, the meson correlation function can be calculated through the generating functional 
$\mathcal{W}[J]=\ln\mathcal{Z}[J]$. We have
\begin{equation}\label{correlation2}
{\cal D}_{\rm M}(X,X^\prime)=\left.\frac{\delta{\cal W}[J] }{\delta J(X)\delta J^\dagger(X^\prime)}\right|_{J=J^\dagger=0}.
\end{equation}
The above formalism is valid for charged mesons (complex scalar bosons). For a neutral meson, i.e., $\phi^\dagger=\phi$, only the source term $J\phi$ is needed and we have
\begin{equation}\label{correlation0}
{\cal D}_{\rm M}(X,X^\prime)=\left.\frac{\delta{\cal W}[J] }{\delta J(X)\delta J(X^\prime)}\right|_{J=0}.
\end{equation}	
	
Now we adopt a chiral effective model of QCD, the NJL model. To study the pion properties, it is sufficient to consider the two-flavor case. The Lagrangian density of the two-flavor NJL model  is given by
\begin{equation}
\mathcal{L}_{\mathrm{NJL}}=\bar{\psi}\left(i \slashed{\partial}-m_{0}\right) \psi+g\left[(\bar{\psi} \psi)^{2}+\left(\bar{\psi} i \gamma_{5} \boldsymbol{\tau}  \psi\right)^{2}\right],
\end{equation}
where $\psi=({\rm u},{\rm d})^{\rm T}$ represents the two-flavor quark field, $m_0$ is the current quark mass, $g$ is the coupling constant of the four-fermion interaction, and $\tau_{\rm i}$~$({\rm i}=1,2,3)$ are the Pauli matrices in the flavor space. In the presence of an external electromagnetic field, the normal derivative $\partial_\mu$ is replaced by the covariant one $D_\mu=\partial_{\mu}-i Q A_{\mu}$, where the quark charge matrix reads
$Q=\text{diag}(Q_{\rm u},Q_{\rm d})$ in the flavor space, with $Q_{\rm u}=2{\rm e}/3$, $Q_{\rm d}=-{\rm e}/3$ and ${\rm e}$ being the elementary charge.  In this work, we consider a constant magnetic field with strength $B$ along the $z$-direction. Thus we choose $A_0=0$ and the vector potential ${\bf A}$ satisfies the equation
$\nabla\times{\bf A}=B \hat{z}$.  The general solution for the vector potential can be expressed as ${\bf A}={\bf A}_\perp+{\bf A}^\prime$, where ${\bf A}^\prime$ has a curl of zero, $\nabla\times{\bf A}^\prime=0$.		
The rotational part ${\bf A}_\perp$ is chosen as
\begin{equation}
\label{gauge}
{\bf A}_\perp=-(1+\xi)\frac{By}{2}\hat{x}+(1-\xi)\frac{Bx}{2}\hat{y}.
\end{equation}
Here the parameter $\xi$ is an arbitrary real number. The symmetric gauge and the Landau gauge correspond to $\xi=0$ and $\xi=\pm 1$, respectively.
	
The partition function of the NJL model  is given by (\ref{partitionF}) with ${\cal L}\rightarrow{\cal L}_{\rm NJL}$. To study charged pions, we introduce the source term $J_+\phi+J_-\phi^\dagger$, where
\begin{eqnarray}
\phi^{\phantom{\dag}}=\bar{\psi}\Gamma_+\psi=\sqrt{2}\bar{\rm u}i\gamma_5{\rm d},\ \ \ \ \ \ \ 
\phi^\dagger=\bar{\psi}\Gamma_-\psi=\sqrt{2}\bar{\rm d}i\gamma_5{\rm u},
\end{eqnarray}
with 
\begin{equation}
\Gamma_\pm\equiv i\gamma_5\frac{\tau_1\pm i\tau_2}{\sqrt{2}}.
\end{equation}
For the neutral pion, the source term is $J\phi$, with $\phi=\bar{\psi}\Gamma_3\psi$ and $\Gamma_3=i\gamma_5\tau_3$.  The four-fermion interaction can be decoupled by applying the 
Hubbard-Stratonovich transformation which also introduces auxiliary meson fields $\sigma(X)$ and $\bm{\pi}(X)$. Integrating out the quark fields, we obtain 
\begin{equation}\label{partitionF1}
{\cal Z}[J]=\int {\cal D}\sigma {\cal D}\bm{\pi} \exp{\Big\{-{\cal S}_J[\sigma,\bm{\pi}]\Big\}}\ ,
\end{equation}
where the action reads
\begin{equation}\label{action}
\mathcal{S}_J\left[\sigma,\boldsymbol{\pi} \right]=\int\mathrm{d}^4X\frac{\sigma^2(X)+\bm{\pi}^2(X)}{4g}-\tilde{\rm Tr}\ln  \boldsymbol{\mathrm{G}}_J^{-1}\left(X,X^\prime\right).
\end{equation}
Here the trace $\tilde{\rm Tr}$ is taken over the coordinate, color, flavor, and spin spaces. To study charged pions, we introduce the sources $J_+$ and $J_-$. The inverse of the quark propagator is given by
\begin{eqnarray}
\boldsymbol{\mathrm{G}}_J^{-1}(X,X^\prime)&=&\Big[i\slashed{D}-m_0-\sigma(X)-i\gamma_5\bm{\tau}\cdot\bm{\pi}(X)
+J_+(X)\Gamma_++J_-(X)\Gamma_-\Big]\ \delta^{(4)}\left(X-X^\prime\right).
\end{eqnarray}
Assuming that the external sources are small, we can expand the generating functional ${\cal W}[J]=\ln{\cal Z}[J]$ in powers of the external sources, ${\cal W}[J]={\cal W}^{(0)}+{\cal W}^{(1)}[J]+{\cal W}^{(2)}[J]+\cdots$. The 
correlation function for charged pions, ${\cal D}_{\rm c}(X,X^\prime)$, is related to the second order term ${\cal W}^{(2)}[J]$. We have
\begin{eqnarray}
{\cal W}^{(2)}[J]=\int {\rm d}^4X\int {\rm d}^4X^\prime J_+(X){\cal D}_{\rm c}(X,X^\prime)J_-(X^\prime).
\end{eqnarray}
To study the neutral pion, we introduce a single source $J$. The inverse of the quark propagator is given by
\begin{eqnarray}
\boldsymbol{\mathrm{G}}_J^{-1}(X,X^\prime)=\Big[i\slashed{D}-m_0-\sigma(X)-i\gamma_5\bm{\tau}\cdot\bm{\pi}(X)+J(X)\Gamma_3\Big]\ \delta^{(4)}\left(X-X'\right).
\end{eqnarray}
The correlation function for the neutral pion, ${\cal D}_{\rm n}(X,X^\prime)$, is related to the second order expansion ${\cal W}^{(2)}[J]$,
\begin{eqnarray}
{\cal W}^{(2)}[J]=\frac{1}{2}\int {\rm d}^4X\int {\rm d}^4X^\prime J(X){\cal D}_{\rm n}(X,X^\prime)J(X^\prime).
\end{eqnarray}

To evaluate the partition function ${\cal Z}[J]$ and the generating functional ${\cal W}[J]$, we use the mean-field approximation (MFA). In this approximation, the quantum fields $\sigma(X)$ and $\bm{\pi}(X)$ are replaced by their classical fields. In the path integral formalism, this is equivalent to replacing the auxiliary fields $\sigma(X)$ and $\bm{\pi}(X)$ with their saddle point values (SPVs), $\sigma_{\rm sp}(X)$ and $\bm{\pi}_{\rm sp}(X)$.  The partition function is now approximated as
\begin{equation}
{\cal Z}[J]\simeq \exp{\Big\{-{\cal S}_J[\sigma_{\rm sp},\bm{\pi}_{\rm sp}]\Big\}}\ .
\end{equation}
The SPVs should be determined by the extreme condition
\begin{equation}
\frac{\delta{\cal S}_J[\sigma_{\rm sp},\bm{\pi}_{\rm sp}]}{\delta\sigma_{\rm sp}}=0,\ \ \ \ \frac{\delta{\cal S}_J[\sigma_{\rm sp},\bm{\pi}_{\rm sp}]}{\delta\bm{\pi}_{\rm sp}}=0.
\end{equation}
In the absence of external sources, the SPVs are static and homogeneous. We set $\sigma_{\rm sp}(X)=\upsilon$, $\bm{\pi}_{\rm sp}(X)={\bf 0}$.
Here $\upsilon=-2g\langle\bar{\psi}\psi\rangle$ contributes to the effective quark mass. However, in the presence of external sources, the SPVs may not be static and homogeneous.  To study the pion correlation functions, we are interested in the response to infinitesimal external sources. We expect that the induced perturbations to the SPVs are also infinitesimal. Therefore, the SPVs can be expressed as
\begin{equation}
\sigma_{\rm sp}(X)=\upsilon+U(X),\ \ \ \ \bm{\pi}_{\rm sp}(X)={\bf 0}+{\bf V}(X),
\end{equation}
where $U(X)$ and ${\bf V}(X)=(V_1,V_2,V_3)$ are infinitesimal perturbations induced by the external sources.
	
In MFA, the generating functional ${\cal W}[J]$ is simply given by
\begin{eqnarray}
\mathcal{W}_{\mathrm{MF}}[J]=-\mathcal{S}_J[\sigma_{\rm sp},\bm{\pi}_{\rm sp}].
\end{eqnarray}
For infinitesimal external sources, it becomes
\begin{eqnarray}
{\cal W}_{\rm MF}[J]=-\frac{1}{4g}\int\mathrm{d}^4X\left\{\left[\upsilon+U(X)\right]^2+2V_+(X)V_-(X)+V_3^2(X)\right\}+ \tilde{\rm Tr}\ln \left[ {\cal G}^{-1}\left(X,X^\prime\right)-\Sigma(X)\delta^{(4)}(X,X^\prime)\right],
\end{eqnarray}
with the notation
\begin{equation}
V_\pm(X)\equiv \frac{V_1(X)\mp i V_2(X)}{\sqrt{2}}.
\end{equation}
Here, the inverse of the quark propagator in MFA reads
\begin{equation}
\mathcal{{G}}^{-1}(X,X^\prime)=\left(i\slashed{D}-M\right)\delta^{(4)}\left(X-X^\prime\right),
\end{equation}
with $M=m_0+\upsilon$ being the effective quark mass. For charged pions, the $J$-dependent part $\Sigma$ is defined as
\begin{eqnarray}
\Sigma(X)=U(X)+\left[V_+(X)-J_+(X)\right]\Gamma_++\left[V_-(X)-J_-(X)\right]\Gamma_-+V_3(X)\Gamma_3.
\end{eqnarray}	
For the neutral pion,  it is
\begin{equation}
\Sigma(X)=U(X)+V_+(X)\Gamma_++V_-(X)\Gamma_-+\left[V_3(X)-J(X)\right]\Gamma_3.
\end{equation}		
	
Using the derivative expansion
\begin{eqnarray}
 \tilde{\rm Tr}\ln \left[ {\cal G}^{-1}\left(X,X^\prime\right)-\Sigma(X)\delta^{(4)}(X,X^\prime)\right]&=& \tilde{\rm Tr}\ln \left[ {\cal G}^{-1}\left(X,X^\prime\right)\right]-\int {\rm d}^4X\ {\rm Tr}\left[{\cal G}(X,X)\Sigma(X)\right]\nonumber\\
&&-\ \frac{1}{2}\int {\rm d}^4X\int {\rm d}^4X^\prime\ {\rm Tr}\left[{\cal G}(X,X^\prime)\Sigma(X^\prime){\cal G}(X^\prime,X)\Sigma(X)\right]+\cdots,
\end{eqnarray}	
we can expand the generating functional ${\cal W}_{\rm MF}[J]$ in powers of the external sources and the induced perturbations,
\begin{eqnarray}
{\cal W}_{\rm MF}[J]={\cal W}_{\rm MF}^{(0)}+{\cal W}_{\rm MF}^{(1)}[J]+{\cal W}_{\rm MF}^{(2)}[J]+\cdots.
\end{eqnarray}	 
Note that the trace over the coordinates has been taken in the expansion and the trace ${\rm Tr}$ is now over the color, flavor, and spin spaces. The extreme of the zeroth order term, $\partial\mathcal{W}_{\mathrm{MF}}^{(0)}/\partial \upsilon=0$, gives rise to the gap equation in a constant magnetic field. Using the vacuum regularization scheme~\cite{Gusynin:1995nb,Avancini:2015ady, Cao:2015xja,Cao:2014uva} and introducing a proper-time variable $s$, we can express the gap equation as
\begin{equation}\label{gapEq}
\frac{M-m_0}{2g}=\frac{N_cM^3}{\pi^2}\left[\frac{\Lambda}{M}\sqrt{1+\frac{\Lambda^2}{M^2}}-\mathrm{arctanh}\left(\frac{\Lambda}{\sqrt{M^2+\Lambda^2}}\right) \right]+\frac{N_cM}{4\pi^2}\sum_{\rm f=u,d}\int_{0}^{\infty} \mathrm{d}s \frac{e^{-sM^2}}{s^2}\left[\frac{\mathcal{B}_{\rm f}s}{\tanh(\mathcal{B}_{\rm f}s)}-1\right].
\end{equation}
Here, $\Lambda$ is the three-momentum cutoff in vacuum, $N_c=3$ is the number of the color degrees of freedom, and $\mathcal{B}_{\rm f}\equiv Q_{\rm f}B$. We note that this form of the gap equation does not depend on the gauge for the vector potential ${\bf A}$. It is easy to check that the first order term  $\mathcal{W}_{\mathrm{MF}}^{(1)}[J]$ vanishes.

The pion correlation functions can be extracted from the second order term ${\cal W}_{\rm MF}^{(2)}[J]$. To evaluate this term, we need to know the expression of the quark propagator ${\cal G}(X,X^\prime)$. A direct Fourier transformation to the momentum space is impossible because of the lack of translational invariance. In the flavor space, it is diagonal and can be written as
\begin{eqnarray}
{\cal G}(X,X^\prime)=\left(\begin{array}{cc}
{\cal G}_{\rm u}(X,X^\prime) & 0\\
0& {\cal G}_{\rm d}(X,X^\prime)
\end{array}\right).
\end{eqnarray}	 
According to Schwinger's proper-time method~\cite{Schwinger:1951nm}, the propagator of each flavor, ${\cal G}_{\rm f}(X,X^\prime)$ (${\rm f=u,d}$), can be decomposed as
\begin{eqnarray}
{\cal G}_{\rm f}(X,X^\prime)=e^{i\Phi_{\rm f}\left(X,X'\right)}\bar{\cal G}_{\rm f}\left(X-X^\prime\right),
\end{eqnarray}	 
where the Schwinger phase reads 
\begin{eqnarray}
\Phi_{\rm f}\left(X,X^\prime\right)=Q_{\rm f}\int_{X^\prime}^X A_\mu dX^\mu=Q_{\rm f}\int_{{\bf r}^\prime}^{\bf r} {\bf A}\cdot {\rm d}{\bf r},
\end{eqnarray}	 
in which the integral is calculated along the straight line. While the Schwinger phase is explicitly gauge dependent and breaks the translation invariance, the remaining part $\bar{\cal G}_{\rm f}\left(X-X^\prime\right)$ is translation invariant and does not depend on the gauge for ${\bf A}$. It is convenient to define the Fourier transformation of $\bar{\cal G}_{\rm f}\left(X-X^\prime\right)$ as
\begin{eqnarray}
\bar{\cal G}_{\rm f}(X-X^\prime)=\int\frac{{\rm d}^4K}{(2\pi)^4}e^{-iK\cdot(X-X^\prime)} \bar{\cal G}_{\rm f}(K),
\end{eqnarray}	 
where we work in the Euclidean space and $K\equiv({\bf K},K_4)$. In our convention, $K\cdot X\equiv K_4\tau-{\bf K}\cdot {\bf r}$.
The momentum-space version is given by
\begin{equation}\label{quark_pro}
\bar{\mathcal{G}}_{\text{f}}(K)=\int_{0}^{\infty} d s~ e^{ -s\left(M^{2}+\mathbf{K}_{\parallel}^{2}+\mathbf{K}_{\perp}^{2} \frac{\tanh (\mathcal{B}_{\rm f} s)}{\mathcal{B}_{\rm f} s}\right)}\left[M-\slashed{K}+i\left(K_{2} \gamma_{1}-K_{1} \gamma_{2}\right) \tanh (\mathcal{B}_{\rm f} s)\right]\Big[1+i \gamma_{1} \gamma_{2} \tanh (\mathcal{B}_{\rm f} s)\Big],
\end{equation}
where we use the notations $\mathbf{K}_{\perp}=(K_1,K_2)$ and $\mathbf{K}_{\parallel}=(K_3,K_4)$ here and in the following.

%%%%%%%%%%%%%%%%%%%%%%%%%%%%%%%%%%%%%%%%%%%%%%%%%%%%%%%%%%%%%%%%%%%%%
\section{The Neutral Pion}\label{sec3}
%%%%%%%%%%%%%%%%%%%%%%%%%%%%%%%%%%%%%%%%%%%%%%%%%%%%%%%%%%%%%%%%%%%%%%
For the neutral pion, we can show that the induced perturbations $U(X)$ and $V_\pm(X)$ do not couple to the source $J(X)$ in the second order term  ${\cal W}_{\rm MF}^{(2)}[J]$, by completing the trace in the flavor space or 
the spin space. Therefore, these induced perturbations should be of order $O(J^2)$ and can be neglected. The relevant terms can be written as
\begin{eqnarray}
{\cal W}_{\rm MF}^{(2)}[J]=-\frac{1}{2}\int {\rm d}^4X\int {\rm d}^4X^\prime 
\left(\begin{array}{cc} J(X) & V_3(X)\end{array}\right)
\left(\begin{array}{cc} \Pi(X,X^\prime) & -\Pi(X,X^\prime)\\ -\Pi(X,X^\prime)& \frac{1}{2g}\delta^{(4)}(X-X^\prime)+\Pi(X,X^\prime)\end{array}\right)
\left(\begin{array}{c} J(X^\prime) \\ V_3(X^\prime)\end{array}\right)
\end{eqnarray}
Here the polarization function for the neutral pion is defined as
\begin{eqnarray}
\Pi(X,X^\prime)=\sum_{\rm f=u,d}{\rm Tr}\left[{\cal G}_{\rm f}(X,X^\prime)i\gamma_5{\cal G}_{\rm f}(X^\prime,X)i\gamma_5\right]
=\sum_{\rm f=u,d}{\rm Tr}\left[\bar{\cal G}_{\rm f}(X-X^\prime)i\gamma_5\bar{\cal G}_{\rm f}(X^\prime-X)i\gamma_5\right].
\end{eqnarray}	 
Here we see that the Schwinger phase of each flavor cancels exactly, in accordance with the charge neutrality of the neutral pion. Performing the Fourier transformations
\begin{eqnarray}
J(X)=\int\frac{{\rm d}^4K}{(2\pi)^4}e^{-iK\cdot X} J(K),\ \ \ \ \ V_3(X)=\int\frac{{\rm d}^4K}{(2\pi)^4}e^{-iK\cdot X} V_3(K),
\end{eqnarray}	 
we obtain
\begin{eqnarray}\label{NP-W}
{\cal W}_{\rm MF}^{(2)}[J]=-\frac{1}{2}\int\frac{{\rm d}^4K}{(2\pi)^4}
\left(\begin{array}{cc} J(-K) & V_3(-K)\end{array}\right)
\left(\begin{array}{cc} \Pi(K) & -\Pi(K)\\ -\Pi(K)& \frac{1}{2g}+\Pi(K)\end{array}\right)
\left(\begin{array}{c} J(K) \\ V_3(K)\end{array}\right),
\end{eqnarray}
where the momentum-space version of the polarization function reads
\begin{eqnarray}
\Pi(K)=\sum_{\rm f=u,d}\int\frac{{\rm d}^4P}{(2\pi)^4}{\rm Tr}\left[\bar{\cal G}_{\rm f}(P)i\gamma_5\bar{\cal G}_{\rm f}(P+K)i\gamma_5\right].
\end{eqnarray}	 
The induced perturbation $V_3(K)$ should be solved by using the extreme condition. Up to order $O(J)$, it is determined by the extreme of ${\cal W}_{\rm MF}^{(2)}$. Using the fact $\Pi(-K)=\Pi(K)$, we find
\begin{eqnarray}
V_3(K)=\frac{\Pi(K)}{\frac{1}{2g}+\Pi(K)}J(K)+O(J^2).
\end{eqnarray}	 
Substituting this solution into (\ref{NP-W}), we finally obtain
\begin{eqnarray}
{\cal W}_{\rm MF}^{(2)}[J]=\frac{1}{2}\int\frac{{\rm d}^4K}{(2\pi)^4}J(-K){\cal D}_{\rm n}(K)J(K),
\end{eqnarray}
where the correlation function of the neutral pion is given by
\begin{eqnarray}
{\cal D}_{\rm n}(K)=-\frac{\Pi(K)}{1+2g\Pi(K)}.
\end{eqnarray}
Its coordinate-space version is obtained via the Fourier transformation
\begin{eqnarray}
{\cal D}_{\rm n}(X,X^\prime)=\int\frac{{\rm d}^4K}{(2\pi)^4}e^{-iK\cdot(X-X^\prime)} {\cal D}_{\rm n}(K).
\end{eqnarray}

The polarization function $\Pi(K)$ can be evaluated as
\begin{eqnarray}
\Pi(K)&=&-4N_c\sum_{\rm f=u,d}\int\frac{\mathrm{d}^4P}{(2\pi)^4}\int_{0}^{\infty}\mathrm{d}s\int_{0}^{\infty}\mathrm{d}t~
e^{-s\left[M^2+\mathbf{P}_\parallel^2+\mathbf{P}_\perp^2\frac{\tanh(\mathcal{B}_{\rm f}s)}{\mathcal{B}_{\rm f}s}\right]}
e^ {-t\left[M^2+\mathbf{(K+P)}_\parallel^2+\mathbf{(K+P)}_\perp^2\frac{\tanh(\mathcal{B}_{\rm f}t)}{\mathcal{B}_{\rm f}t}\right]}\nonumber\\
&&\times\Bigg\{\text{sech}^2(\mathcal{B}_{\rm f}s)\text{sech}^2(\mathcal{B}_{\rm f}t)\mathbf{P}_\perp\cdot\mathbf{(K+P)}_\perp+\left[M^2+\mathbf{P}_\parallel\cdot\mathbf{(K+P)}_\parallel\right]
\Big[1+\tanh(\mathcal{B}_{\rm f}s)\tanh(\mathcal{B}_{\rm f}t)\Big]\Bigg\}.
\end{eqnarray}
In accordance with the gap equation, here we also use the vacuum regularization scheme~\cite{Cao:2015xja,Cao:2014uva,Gusynin:1995nb,Avancini:2015ady} to regularize the ultraviolet divergence. The polarization function 
is decomposed into a vacuum term and a $B$-dependent term,
\begin{equation}\label{polarizationp0}
\Pi(K)=\Pi_{V}(K)+\Pi_B(K).
\end{equation}
The divergence in the vacuum term is regularized by using the same three-momentum cutoff $\Lambda$ as in the gap equation Eq.~(\ref{gapEq}). We write
\begin{equation}\label{Re-Vacuum}
\Pi_V(K)=-8N_c\int\frac{\mathrm{d}^4P}{(2\pi)^4}\frac{M^2+P\cdot(P+K)}{\left[M^2+(P\!+K)^2\right]\left(M^2+\!P^2\right)}\Theta(\Lambda-|{\bf P}|).
\end{equation}
Here the notation $P\cdot K\equiv {\bf P}\cdot{\bf K}+P_4K_4$ is used.
The $B$-dependent term $\Pi_B(K)=\Pi(K)-\lim\limits_{B\rightarrow0}\Pi(K)$ is finite and characterizes the effects induced by the magnetic field.

The space-integrated correlation function of the neutral pion, ${\cal P}_{\rm n}(\tau)$, which is only a function of the imaginary time $\tau$, can be evaluated as
\begin{eqnarray}
{\cal P}_{\rm n}(\tau)&=&\int{\rm d}^3{\bf r}{\cal D}_{\rm n}(\tau,{\bf r};\tau^\prime=0,{\bf r}^\prime={\bf 0})\nonumber\\
&=&\int\mathrm{d}^3\mathbf{r}\int\frac{\mathrm{d}^4K}{(2\pi)^4}e^{-i\left(K_4\tau-\mathbf{K}\cdot\mathbf{r}\right)}{\cal D}_{\rm n}(K)\nonumber\\
&=&\int_{-\infty}^{\infty}\frac{\mathrm{d}K_4}{2\pi}e^{-iK_4\tau}{\cal D}_{\rm n}(K_4,{\bf K}={\bf 0})\nonumber\\
&=&\int_{0}^{\infty}\frac{\mathrm{d}K_4}{\pi}\cos(K_4\tau){\cal D}_{\rm n}(K_4,{\bf K}={\bf 0}).
\end{eqnarray}
In the last line, we have used the fact that the polarization function $\Pi(K)$ is even in $K$. Here we see that the integration over ${\bf r}$ forces the momentum ${\bf K}$ to be zero. For vanishing $\mathbf{K}$, the polarization function $\Pi(K)$ can be simplified. Completing the integral over the four-momentum $P$ and substituting the proper-time variables with $s\rightarrow (1+u)s/2$ and $t\rightarrow (1-u)s/2$, we obtain
\begin{equation}
\Pi(K_4,{\bf K}={\bf 0})=\frac{N_c}{8\pi^2}\sum_{\rm f=u,d}\int_{0}^{\infty}\mathrm{d}s\int_{-1}^{1}\mathrm{d}u \ e^{-\left(\frac{1\!-\!u^2}{4}K_4^2\!+\!M^2\right)s}\left\{\left[\frac{1\!-\!u^2}{4}\!K_4^2-\!\frac{1}{s}\!-\!M^2\right]
\mathcal{B}_{\rm f}\coth(\mathcal{B}_{\rm f}s)-\frac{\mathcal{B}_{\rm f}^2}{\sinh^2(\mathcal{B}_{\rm f}s)} \right\}.
\end{equation}
Meanwhile, the vacuum term in Eq.~(\ref{Re-Vacuum}) becomes 
\begin{equation}
\Pi_V(K_4,{\bf K}={\bf 0})=-\frac{8N_c}{\pi^2}\int_{0}^{\Lambda}\mathrm{d}|{\bf P}|\frac{{\bf P}^2\sqrt{{\bf P}^2+M^2}}{4({\bf P}^2+M^2)+K_4^2}.
\end{equation}

%---------------------------------------------------------------------
\begin{figure}[H]
	\centering
	\includegraphics[width=0.38\textwidth]{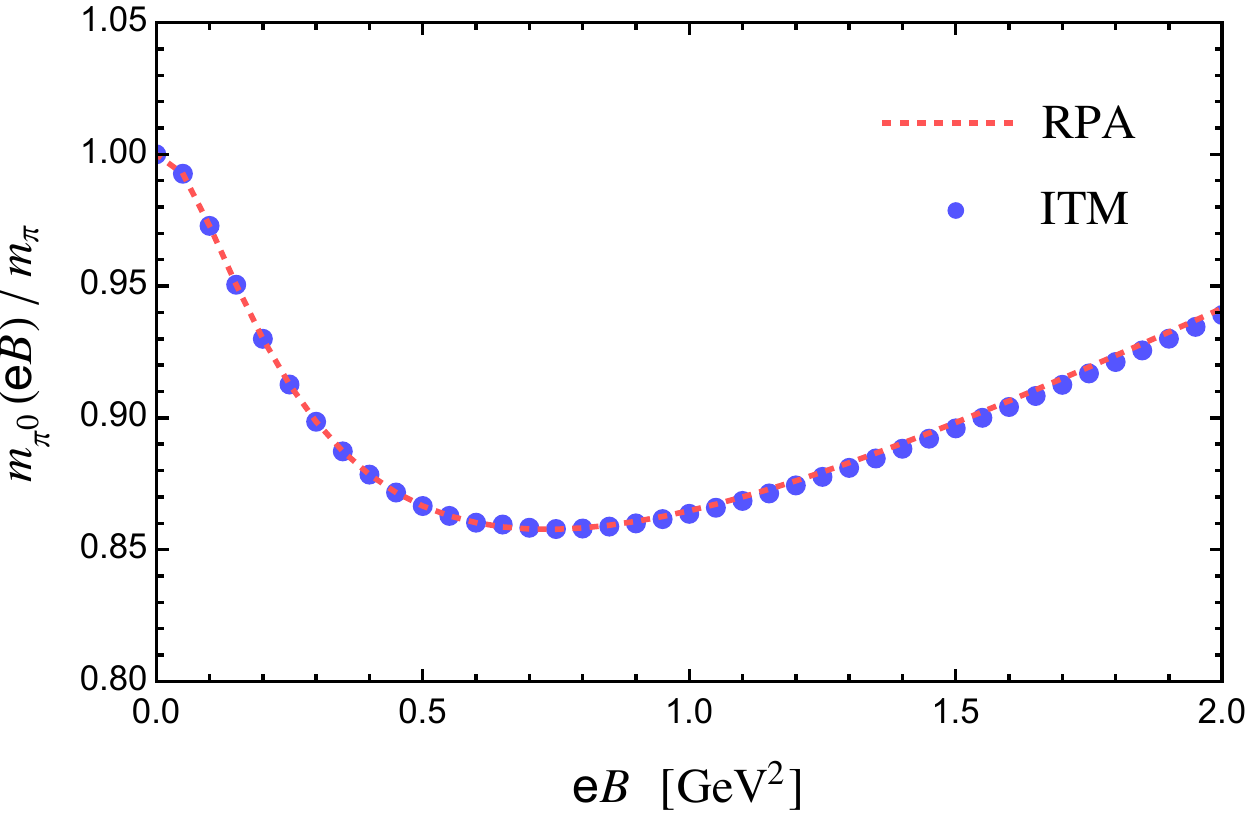}
	\caption{The mass of the neutral pion, $m_{\pi^0}({\rm e}B)$, normalized by the vacuum pion mass $m_\pi=0.134$~GeV, as a function of ${\rm e}B$. The blue dots are our results extracted from the correlation function in the NJL model using the imaginary-time method. The result from the momentum-space RPA (red dotted line) 
		is presented for comparison.}
	\label{neutral_pion}
\end{figure}
%---------------------------------------------------------------------
	
%---------------------------------------------------------------------
\begin{figure}[H]
	\centering
	\includegraphics[width=0.38\textwidth]{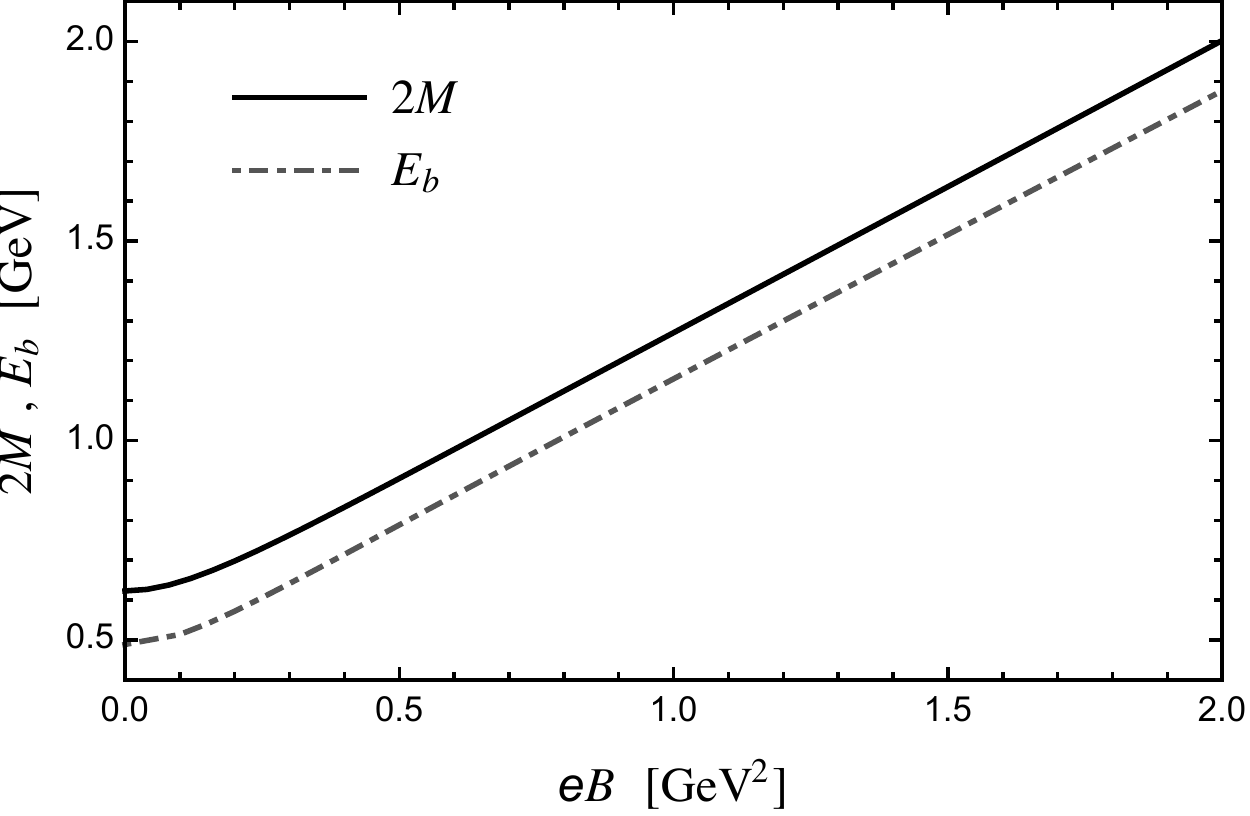}
	\caption{The effective quark mass $M$ and the binding energy $E_{\rm b}$ as a function of ${\rm e}B$.}
	\label{mass_and_potential}
\end{figure}
%---------------------------------------------------------------------

Now we present our numerical results. For the two-flavor NJL model, we use the the parameter set $g=4.93~{\rm GeV}^{-2}$, $\Lambda=0.653~{\rm GeV}$, and $m_0=5~{\rm MeV}$, determined by fitting the pion mass $m_\pi=0.134~{\rm GeV}$, the pion decay constant $f_\pi=93~{\rm MeV}$, and the quark condensate $\left<\bar{\rm u}{\rm u}\right>=-(0.25~{\rm GeV})^3 $ in the vacuum at vanishing external magnetic field~\cite{Zhuang:1994dw}.  We calculate the the neutral pion mass $m_{\pi^0}({\rm e}B)$ by fitting the large-$\tau$ behavior ${\cal P}_{\rm n}\sim \exp{(-\tau m_{\pi^0})}$. The numerical result is shown in Fig.~\ref{neutral_pion}.

As a comparison, we also show the pole mass obtained from the momentum-space RPA~\cite{Cao:2019res}, i.e., solution of the pole equation in the Minkowski space,
\begin{equation}
1+2g\Pi(K_4\rightarrow -iK_0,{\bf K}={\bf 0})=0.
\end{equation}
The two results are in perfect agreement as we expected. The agreement can be analytically proven as follows. Let us consider the following integral
\begin{eqnarray}
{\cal I}(\tau)=\int_{-\infty}^{\infty}\frac{\mathrm{d}K_4}{2\pi}e^{-iK_4\tau}F(K_4),
\end{eqnarray}
where $F(K_4)$ is an arbitrary function. Rotating to the Minkowski space by substituting $K_4\rightarrow -iK_0$, we obtain
\begin{eqnarray}
{\cal I}(\tau)=\int_{-i\infty}^{i\infty}\frac{\mathrm{d}K_0}{2\pi i}e^{-K_0\tau}f(K_0),
\end{eqnarray}
where $f(K_0)\equiv F(-iK_0)$. To proceed, we suppose that $F(K_4)$ has no real poles for $K_4$, i.e., $f(K_0)$ has no poles on the imaginary axis of the complex $K_0$-plane.
For $\tau>0$, we can close the integral path along the imaginary axis with a semicircle at infinity in the right half plane, obtaining
\begin{eqnarray}
{\cal I}(\tau)=\oint_C \frac{\mathrm{d}K_0}{2\pi i}e^{-K_0\tau}f(K_0)=\sum_l {\rm Res}f(E_l) e^{-E_l\tau},
\end{eqnarray}
where $E_l$ ($l=0,1,2,...$) are the poles of $f(K_0)$ in the right half plane and ${\rm Res}f(E_l)$ are the corresponding residues. Assuming that the pole $E_0$ with minimal real part is real, we find that for $\tau\rightarrow\infty$, ${\cal I}(\tau)\sim e^{-E_0\tau} $. Thus the obtained mass $E_0$ is exactly the same as the pole mass solved from the pole equation.

While the neutral pion mass from the above NJL model calculation shows a non-monotonic behavior, i.e., it turns to increase at large magnetic field (${\rm e}B\textgreater0.8$GeV$^2$), recent lattice QCD calculations indicate a monotonic decrease~\cite{Luschevskaya:2015cko,Bali:2017ian,Ding:2020jui,Ding:2020hxw}. This could be understood from the binding energy of the neutral pion, $E_{\rm b}=2M-m_{\pi^0}$. The numerical result from the NJL model is demonstrated in Fig.~\ref{mass_and_potential}. If the monotonic decrease observed in lattice QCD calculations is true, it indicates that the binding energy $E_{\rm b}$ is underestimated or the magnetic catalysis is overestimated in the NJL model. We may need to improve the NJL model by adopting a nonlocal interaction~\cite{GomezDumm:2017jij} or an $\text{e}B$-dependent coupling constant~\cite{Avancini:2016fgq}.

%%%%%%%%%%%%%%%%%%%%%%%%%%%%%%%%%%%%%%%%%%%%%%%
\section{Charged Pions}\label{sec4}
%%%%%%%%%%%%%%%%%%%%%%%%%%%%%%%%%%%%%%%%%%%%%%%
For charged pions, we can show that the induced perturbations $U(X)$ and $V_3(X)$ do not couple to the sources $J_\pm(X)$ in the second order term  ${\cal W}_{\rm MF}^{(2)}[J]$, by completing the trace in the flavor space or 
the spin space. Therefore, these induced perturbations should be of order $O(J_\pm^2)$ and can be neglected. The relevant terms can be written as
\begin{eqnarray}
\label{Wmf}
{\cal W}_{\rm MF}^{(2)}[J]=-\int {\rm d}^4X\int {\rm d}^4X^\prime 
\left(\begin{array}{cc} J_+(X) & V_+(X)\end{array}\right)
\left(\begin{array}{cc} \Pi(X,X^\prime) & -\Pi(X,X^\prime)\\ -\Pi(X,X^\prime)& \frac{1}{2g}\delta^{(4)}(X-X^\prime)+\Pi(X,X^\prime)\end{array}\right)
\left(\begin{array}{c} J_-(X^\prime) \\ V_-(X^\prime)\end{array}\right)
\end{eqnarray}
Here the polarization function for charged pions is defined as
\begin{eqnarray}
\Pi(X,X^\prime)=2{\rm Tr}\left[{\cal G}_{\rm u}(X,X^\prime)i\gamma_5{\cal G}_{\rm d}(X^\prime,X)i\gamma_5\right]
=2e^{i\Phi(X,X^\prime)}{\rm Tr}\left[\bar{\cal G}_{\rm u}(X-X^\prime)i\gamma_5\bar{\cal G}_{\rm d}(X^\prime-X)i\gamma_5\right].
\end{eqnarray}	 
The Schwinger phase $\Phi(X,X^\prime)$ can be evaluated as
\begin{eqnarray}
 \Phi(X,X^\prime)=\Phi_{\rm u}(X,X^\prime)+\Phi_{\rm d}(X^\prime,X)=Q_{\rm u}\int_{X^\prime}^X A_\mu dX^\mu+Q_{\rm d}\int^{X^\prime}_X A_\mu dX^\mu={\rm e}\int_{X^\prime}^X A_\mu dX^\mu.
\end{eqnarray}	 
Thus $ \Phi(X,X^\prime)$ is just the Schwinger phase of the composite charged pions.
For any two spacetime points $Z$ and $Z^\prime$, the correlation function of charged pions, $\mathcal{D}_\text{c}\left(Z,Z^\prime\right)$, is given by
\begin{equation}
\label{Dc-definition}
\mathcal{D}_\text{c}\left(Z,Z'\right)=\frac{\delta^2\mathcal{W}^{(2)}_\text{MF}\left[J\right]}{\delta J_+\left(Z\right)\delta J_-\left(Z'\right)}.
\end{equation}

To separate the Schwinger phase, we define the following reduced generating functional
\begin{eqnarray}
\label{Wbar}
\overline{{\cal W}}_{\rm MF}^{(2)}[\bar J]=-\int {\rm d}^4X\int {\rm d}^4X^\prime 
\left(\begin{array}{cc} \bar{J}_+(X) & \bar{V}_+(X)\end{array}\right)
\left(\begin{array}{cc} \bar\Pi(X-X^\prime) & -\bar\Pi(X-X^\prime)\\ -\bar\Pi(X-X^\prime)& \frac{1}{2g}\delta^{(4)}(X-X^\prime)+\bar\Pi(X-X^\prime)\end{array}\right)
\left(\begin{array}{c} \bar{J}_-(X^\prime) \\ \bar{V}_-(X^\prime)\end{array}\right),
\end{eqnarray}
where the reduced polarization function is defined as
\begin{eqnarray}
\bar{\Pi}(X-X^\prime)=2{\rm Tr}\left[\bar{\cal G}_{\rm u}(X-X^\prime)i\gamma_5\bar{\cal G}_{\rm d}(X^\prime-X)i\gamma_5\right],
\end{eqnarray}	
and the recuded sources are given by
\begin{eqnarray}
\bar{J}_+(X)=J_+(X)e^{i\Phi(X,Z^\prime)},\ \ \ \ \bar{V}_+(X)=V_+(X)e^{i\Phi(X,Z^\prime)},\ \ \ \ \bar{J}_-(X^\prime)=J_-(X^\prime)e^{i\Phi(Z^\prime,X^\prime)},\ \ \ \ \bar{V}_-(X^\prime)=V_-(X^\prime)e^{i\Phi(Z^\prime,X^\prime)}.
\end{eqnarray}
Note that here $Z^\prime$ is an arbitrary given spacetime point designed to calculate the correlator $\mathcal{D}_\text{c}\left(Z,Z^\prime\right)$. Expression (\ref{Wbar}) has formal translational invariance.
If we reexpress it in terms of the original sources $J_\pm$ and $V_\pm$, the phase term $\exp\left[i\Phi\left(X,Z^\prime\right)+i\Phi\left(Z^\prime,X^\prime\right)\right]$ appears. This is in contrast to the Schwinger phase term  $\exp\left[i\Phi(X,X^\prime)\right]$ appearing in (\ref{Wmf}). Using this reduced generating functional, we can show that
\begin{equation}
\label{redefinitions}
\mathcal{D}_\text{c}\left(Z,Z^\prime\right)=e^{i\Phi\left(Z,Z^\prime\right)}\frac{\delta^2\overline{\mathcal{W}}^{(2)}_\text{MF}[\bar J]}{\delta \bar J_+\left(Z\right)\delta \bar J_-\left(Z^\prime\right)}.
\end{equation}
The proof of this result is tedious but straightforward. The details are presented in Appendix \ref{Appendix}. We note that the phase term $\exp\left[i\Phi(Z',X^\prime)\right]$ reduces to unity automatically and the phase term $\exp\left[i\Phi(X,Z^\prime)\right]$ reduces to the Schwinger phase $\exp\left[i\Phi(X,X^\prime)\right]$  when we do the variational derivative with respect to the source $\bar J_-$ located at $Z'$. 

Performing the Fourier transformations
\begin{eqnarray}
\bar{J}_\pm(X)=\int\frac{{\rm d}^4K}{(2\pi)^4}e^{-iK\cdot X} \bar{J}_\pm(K),\ \ \ \ \ \bar{V}_\pm(X)=\int\frac{{\rm d}^4K}{(2\pi)^4}e^{-iK\cdot X} \bar{V}_\pm(K),
\end{eqnarray}	 
we obtain
\begin{eqnarray}\label{CP-W}
\overline{{\cal W}}_{\rm MF}^{(2)}[\bar J]=-\int\frac{{\rm d}^4K}{(2\pi)^4}
\left(\begin{array}{cc} \bar{J}_+(-K) & \bar{V}_+(-K)\end{array}\right)
\left(\begin{array}{cc} \bar\Pi(K) & -\bar\Pi(K)\\ -\bar\Pi(K)& \frac{1}{2g}+\bar\Pi(K)\end{array}\right)
\left(\begin{array}{c} \bar{J}_-(K) \\ \bar{V}_-(K)\end{array}\right),
\end{eqnarray}
where the momentum-space version of the reduced polarization function reads
\begin{eqnarray}
\bar\Pi(K)=2\int\frac{{\rm d}^4P}{(2\pi)^4}{\rm Tr}\left[\bar{\cal G}_{\rm u}(P)i\gamma_5\bar{\cal G}_{\rm d}(P+K)i\gamma_5\right].
\end{eqnarray}	 
The induced perturbations $\bar V_\pm (K)$ should be solved by using the extreme condition. Up to order $O(\bar J)$, it can be determined by the extreme of $\overline{{\cal W}}_{\rm MF}^{(2)}$. We obtain
\begin{eqnarray}
\bar{V}_\pm(K)=\frac{\bar\Pi(K)}{\frac{1}{2g}+\bar\Pi(K)}\bar{J}_\pm(K)+O(\bar J^2).
\end{eqnarray}	 
Substituting this solution into (\ref{CP-W}), we finally obtain
\begin{eqnarray}
\overline{{\cal W}}_{\rm MF}^{(2)}[\bar J]=\int\frac{{\rm d}^4K}{(2\pi)^4}\bar{J}_+(-K)\bar{\cal D}_{\rm c}(K)\bar{J}_-(K),
\end{eqnarray}
where
\begin{eqnarray}
\bar{\cal D}_{\rm c}(K)=-\frac{\bar\Pi(K)}{1+2g\bar\Pi(K)}
\end{eqnarray}
can be understood as the reduced correlation function of charged pions in the momentum space without the Schwinger phase.

Therefore, the full correlation function of charged pions based on Eq.~(\ref{redefinitions}) reads
\begin{eqnarray}
{\cal D}_{\rm c}(Z,Z^\prime)=e^{i\Phi(Z,Z^\prime)}\int\frac{{\rm d}^4K}{(2\pi)^4}e^{-iK\cdot(Z-Z^\prime)}\bar{\cal D}_{\rm c}(K)
=-e^{i\Phi(Z,Z^\prime)}\int\frac{{\rm d}^4K}{(2\pi)^4}e^{-iK\cdot(Z-Z^\prime)}\frac{\bar\Pi(K)}{1+2g\bar\Pi(K)}.
\end{eqnarray}
The reduced polarization function $\bar\Pi(K)$ can be evaluated as
\begin{eqnarray}
\bar\Pi(K)&=&-8N_c\int\frac{\mathrm{d}^4P}{(2\pi)^4}\int_{0}^{\infty}\mathrm{d}s\mathrm{d}t~e^{-s\left[M^2+\mathbf{P}_\parallel^2+\mathbf{P}_\perp^2\frac{\tanh(\mathcal{B}_{\rm u}s)}{\mathcal{B}_{\rm u}s}\right]}
e^ {-t\left[M^2+\mathbf{(P+K)}_\parallel^2+\mathbf{(P+K)}_\perp^2\frac{\tanh(\mathcal{B}_{\rm d}t)}{\mathcal{B}_{\rm d}t}\right]}\nonumber\\
&&\times\ \Bigg\{\text{sech}^2(\mathcal{B}_{\rm u}s)\text{sech}^2(\mathcal{B}_{\rm d}t)\mathbf{P}_\perp\cdot\mathbf{(P+K)}_\perp+\left[M^2+\mathbf{P}_\parallel\cdot\mathbf{(P+K)}_\parallel\right]
\Big[1+\tanh(\mathcal{B}_{\rm u}s)\tanh(\mathcal{B}_{\rm d}s)\Big]\Bigg\}.
\end{eqnarray}
Again it is decomposed into a vacuum term and a $B$-dependent term,
\begin{equation}
\bar\Pi(K)=\Pi_{V}(K)+\bar\Pi_B(K).
\end{equation}
The regularized version of the vacuum term $\Pi_V(K)$ is given by Eq. (\ref{Re-Vacuum}).

For the general decomposition of the vector potential, ${\bf A}={\bf A}_\perp+{\bf A}^\prime$, the Schwinger phase $\Phi$ can be evaluated as
\begin{equation}
\Phi\left(\mathbf{r},\mathbf{r}^\prime\right)={\rm e}\int_{\mathbf{r}^\prime}^{\mathbf{r}}{\bf A}\cdot {\rm d}{\bf r}
=-\frac{{\rm e}B}{2}\left[(xy^\prime-x^\prime y)+\xi(xy-x^\prime y^\prime)\right]+{\rm e}\left[\varphi({\bf r})-\varphi({\bf r}^\prime)\right].
\end{equation}
Here we have defined ${\bf A}^\prime=\nabla\varphi$ due to the fact that ${\bf A}^\prime$ has a curl of zero. Then, the space-integrated correlation function of charged pions, ${\cal P}_{\rm c}(\tau)$, which is only a function of the imaginary time $\tau$, is given by
\begin{eqnarray}\label{P-tau-C}
{\cal P}_{\rm c}(\tau)=\int{\rm d}^3{\bf r}\ {\cal D}_{\rm c}(\tau,{\bf r};\tau^\prime=0,{\bf r}^\prime={\bf 0})
=\int\mathrm{d}^3\mathbf{r}\ \left[e^{i\Phi\left(\mathbf{r},\mathbf{0}\right)}\int\frac{\mathrm{d}^4K}{(2\pi)^4}e^{-i\left(K_4\tau-\mathbf{K}\cdot\mathbf{r}\right)}\bar{\cal D}_{\rm c}(K)\right].
\end{eqnarray}
A complete study of the gauge dependence is rather complicated due to the large functional space of $\varphi({\bf r})$. In the following, we first neglect ${\bf A}^\prime$ and focus on the dependence on the parameter $\xi$.  In this case, the integral over ${\bf r}$ can be carried out to obtain
\begin{eqnarray}\label{decay_charged_pion}
{\cal P}_{\rm c}(\tau)
=\int_{0}^{\infty}\frac{\mathrm{d}K_4}{\pi}\cos\left(K_4\tau\right)\int_{0}^{\infty}\frac{\mathrm{d}\mathbf{K}_\perp^2}{{\rm e}B\xi}J_0\left(\frac{\mathbf{K}_\perp^2}{{\rm e}B\xi}\right)\bar{\cal D}_{\rm c}(K_4,\mathbf{K}_\perp,K_3=0),
\end{eqnarray}
Here $J_0(x)$ is the zeroth-order Bessel function of the first kind. Unlike the case for the neutral pion, here the space integration only forces $K_3=0$.  If we work in in the symmetry gauge ($\xi=0$), the transverse momentum is also forced to be zero, because of the fact
\begin{equation}
\lim\limits_{\xi\rightarrow0}\frac{1}{{\rm e}B\xi}J_0\left(\frac{\mathbf{K}_\perp^2}{{\rm e}B\xi}\right)=\delta\left(\mathbf{K}_\perp^2\right)\label{c1}.
\end{equation}
For vanishing $K_3$, the polarization function $\bar\Pi(K)$ can be further simplified by substituting the proper-time variables with $s\rightarrow (1+u)s/2$ and $t\rightarrow (1-u)s/2$. We obtain
\begin{eqnarray}\label{charged_polarization}
\bar\Pi(K_4,\mathbf{K}_\perp,K_3=0)&=&\frac{N_c}{4\pi^2}\int_{0}^{\infty}\!\mathrm{d}s\int_{-1}^{1}\!\mathrm{d}u \ 
e^{-\frac{1-u^2}{4}sK_4^2-sM^2-{\cal I}_1(s,u)\mathbf{K}_\perp^2}{\cal I}_2(s,u)
\Bigg\{\left[\frac{1\!-\!u^2}{4}K_4^2-\frac{1}{s}-M^2\right]\Big[1+\tanh\left(\mathcal{B}_{\rm u}^-s\right)\tanh\left(\mathcal{B}_{\rm d}^+s\right) \Big]\nonumber\\
&&+\ {\cal I}_2(s,u)\text{sech}^2\left(\mathcal{B}_{\rm u}^-s\right)\text{sech}^2\left(\mathcal{B}_{\rm d}^+s\right)\left[{\cal I}_1(s,u)\mathbf{K}_\perp^2-1\right]\Bigg\},
\end{eqnarray}
where $\mathcal{B}_{\rm u}^-=\mathcal{B}_{\rm u}(1-u)/2$ and $\mathcal{B}_{\rm d}^+=\mathcal{B}_{\rm d}(1+u)/2$. The functions ${\cal I}_1(s,u)$ and ${\cal I}_2(s,u)$ are defined as
\begin{eqnarray}
{\cal I}_1(s,u)=\frac{1}{{\mathcal{B}_{\rm u}\coth\left(\mathcal{B}_{\rm u}^-s\right)}+\mathcal{B}_{\rm d}\coth\left(\mathcal{B}_{\rm d}^+s\right)},\ \ \ \ \ \ 
{\cal I}_2(s,u)=\frac{\mathcal{B}_{\rm u}\mathcal{B}_{\rm d}}{\mathcal{B}_{\rm u}\tanh\left(\mathcal{B}_{\rm d}^+s\right)+\mathcal{B}_{\rm d}\tanh\left(\mathcal{B}_{\rm u}^-s\right)}
\end{eqnarray}

%%%%%%%%%%%%%%%%%%%%%%%%%%%%%%%%%%%%%%%%%%%%%%%%%%%%%%%%%%%
\subsubsection{Charged Pions as Point Particles}
%%%%%%%%%%%%%%%%%%%%%%%%%%%%%%%%%%%%%%%%%%%%%%%%%%%%%%%%%%%
Before we present the numerical result of charged pions in the NJL model, it is useful to study the corresponding results in the free Klein-Gordon theory, i.e., consider them as free point particles. The quantized energy levels, i.e., the Landau levels of charged pions, obtained from the Klein-Gordon equation in a constant magnetic field, is given by
\begin{equation}
E^2(n,K_3)=K_3^2+(2n+1){\rm e}B+m_\pi^2,
\end{equation}
with $n=0,1,2,~\dots$ characterizing the Landau levels. Note that these Landau levels are independent of the gauge chosen for the vector potential ${\bf A}$. Therefore, the charged pion mass is given by the energy of the lowest Landau level (LLL), i.e.,
\begin{equation}
m_{\pi^\pm}({\rm e}B)=\min{\{|E(n,K_3)|\}}=\sqrt{m_\pi^2+{\rm e}B}.
\end{equation}
	
\begin{figure}[H]
	\centering
\includegraphics[width=0.4\textwidth]{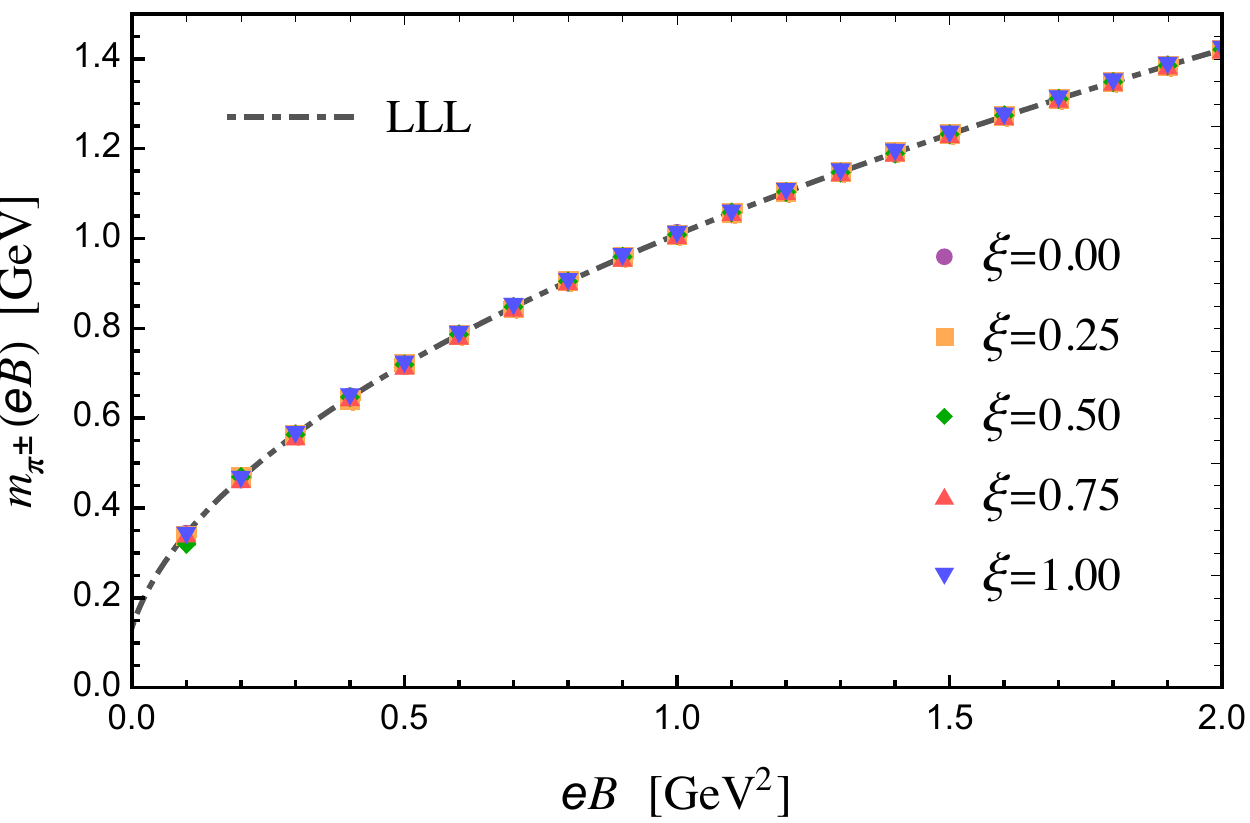}
\caption{The mass of charged pions, $m_{\pi^\pm}({\rm e}B)$, as a function of ${\rm e}B$, extracted from the large-$\tau$ behavior of Eq. (\ref{point_particle_model}) for different 
gauge paramater $\xi$ varying from $0$ to $1$. The analytical result (LLL) is also shown for comparison.}
\label{point_particle}
\end{figure}

On the other hand, we can also determine the charged pion mass from their correlation function by using the imaginary-time method. The Fourier transformation of their reduced correlation function is given by ~\cite{Kuznetsov:2015uca}:
\begin{equation}\label{point_propagator}
\bar{\cal D}_{\rm c}(K)=\int_{0}^{\infty} \mathrm{d}s\ {\rm sech}({\rm e}Bs)\exp{\left\{-s\left[m_\pi^2+\mathbf{K}_{\parallel}^{2}+\mathbf{K}_{\perp}^{2} \frac{\tanh ({\rm e}Bs)}{{\rm e}Bs}\right]\right\}}.
\end{equation}
The space-integrated correlation function can be analytically evaluated as
\begin{eqnarray}\label{point_particle_model}
{\cal P}_{\rm c}(\tau)&=&\int\mathrm{d}^3\mathbf{r}\ \left[e^{i\Phi\left(\mathbf{r},\mathbf{0}\right)}\int\frac{\mathrm{d}^4K}{(2\pi)^4}e^{-i\left(K_4\tau-\mathbf{K}\cdot\mathbf{r}\right)}\bar{\cal D}_{\rm c}(K)\right]\nonumber\\
&=&\frac{1}{2\sqrt{\pi}}\int_{0}^{\infty}\mathrm{d}s\frac{\text{csch}({\rm e}Bs)}{\sqrt{s\left[\xi^2+\coth^2({\rm e}Bs)\right]}}\exp{\left(-sm_\pi^2-\frac{\tau^2}{4s}\right)}.
\end{eqnarray}
In the limit $\tau\rightarrow\infty$, the integral over $s$ is dominated around $s\sim\tau$ due to the term $\exp{[-\tau^2/(4s)]}$. For large $s$, we can use the leading asymptotic behavior 
${\rm csch}({\rm e}Bs)\sim 2\exp(-{\rm e}Bs)$ and $\coth({\rm e}Bs)\sim 1$. The proper-time integral can be carried out analytically, obtaining
\begin{equation}
\mathcal{P}_{\rm c}(\tau)\sim \frac{1}{\sqrt{\xi^2+1}}\frac{1}{\sqrt{m_\pi^2+{\rm e}B}}\exp{\left(-\tau \sqrt{m_\pi^2+{\rm e}B}\right)},\ \ \ \ \ \tau\rightarrow\infty.
\end{equation}
It is clear that even though the prefactor depends on the gauge parameter $\xi$, the mass is always the same. We also calculate the charged pion mass numerically. The results are shown in Fig.~\ref{point_particle}. The numerical results for different gauge parameters are in perfect agreement with the mass determined by the energy of LLL.
	
%---------------------------------------------------------------------
%---------------------------------------------------------------------

\begin{figure}[H]
	\centering
	\includegraphics[width=0.4\textwidth]{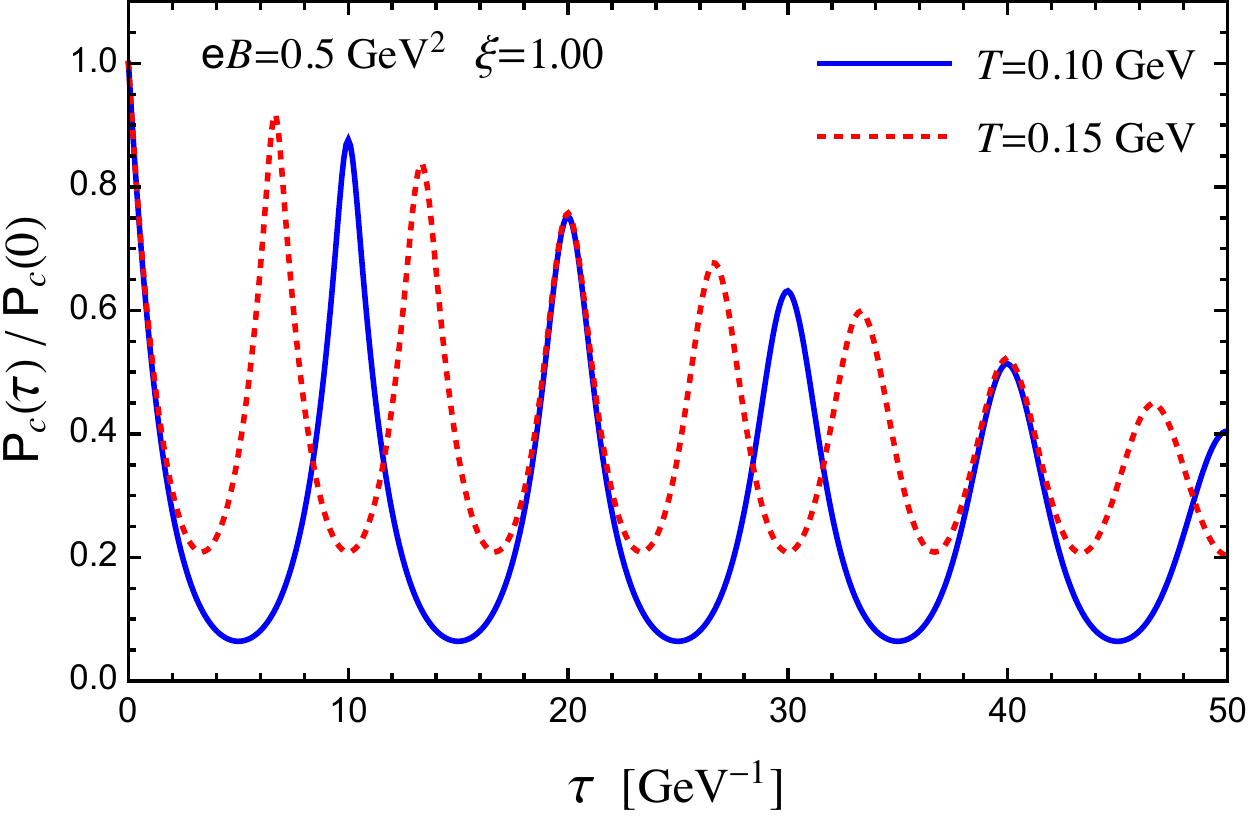}
	\caption{The behavior of the correlation function ${\cal P}_{\rm c}(\tau)$ at finite temperature in the free point particle model.}
	\label{Point_Charged_Thermal_Correlation}
\end{figure}

\begin{figure}[H]
	\centering
	\includegraphics[width=0.4\textwidth]{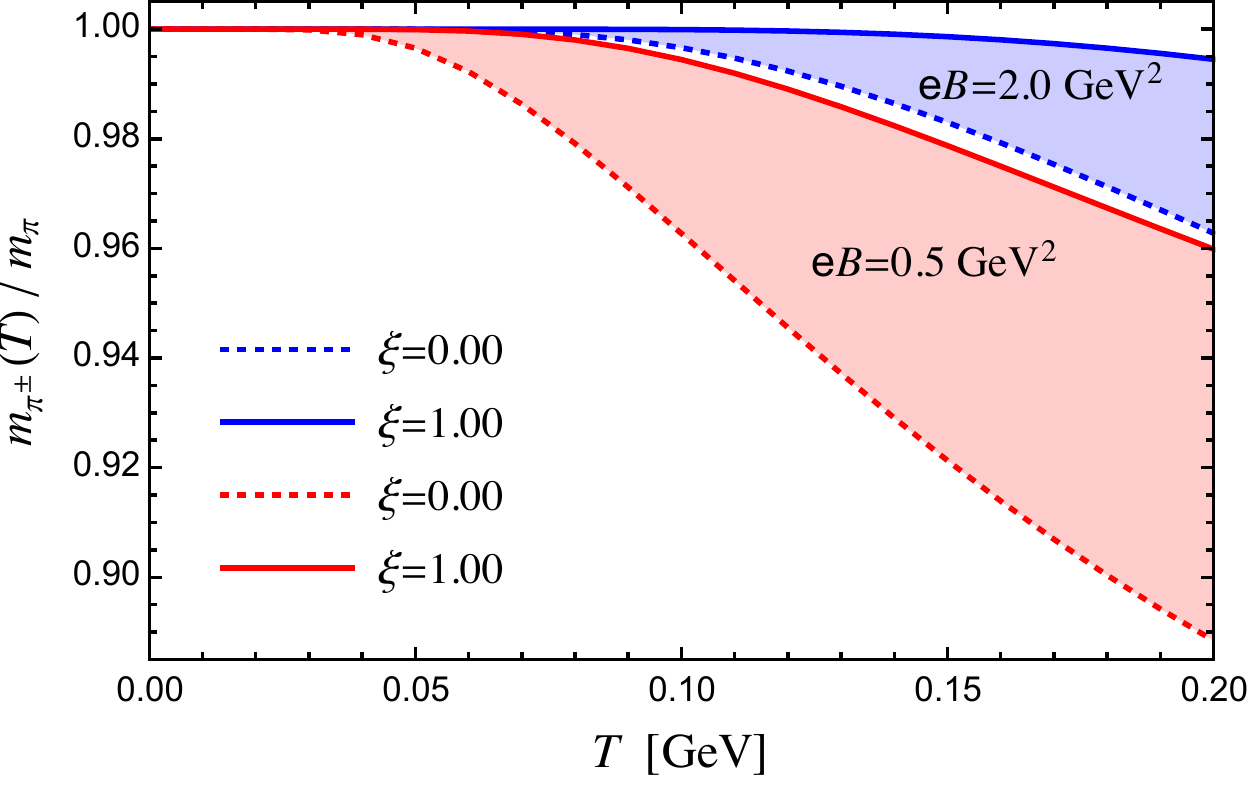}
	\caption{The charged pion mass in the free point particle model as a function of $T$ with the gauge constant $\xi$ varying from $0$ to $1$ for ${\rm e}B=0.5$ $\text{GeV}^2$ (red band) and 
	${\rm e}B=2.0$ $\text{GeV}^2$ (blue band). }
	\label{Thermal}
\end{figure}

It is interesting to discuss whether we can extend the imaginary-time method to finite temperature. At finite temperature $T$, the correlation function can be obtained via the following replacements:
\begin{equation}
K_4\rightarrow 2n\pi T,\ \ \ \ \ \ \ \ \int_{-\infty}^\infty\frac{{\rm d}K_4}{2\pi}\rightarrow T\sum_{n=-\infty}^\infty.
\end{equation}
Accordingly, Eq.~\eqref{point_particle_model} becomes
\begin{equation}\label{point_particle_model_T}
{\cal P}_{\rm c}(\tau)=\frac{1}{2\sqrt{\pi}}\int_{0}^{\infty}\mathrm{d}s\frac{\text{csch}({\rm e}Bs)}{\sqrt{s\left[\xi^2+\coth^2({\rm e}Bs)\right]}}\exp{\left(-sm_\pi^{2}-\frac{\tau^2}{4s}\right)}
\vartheta_3\left(-{i\,\tau\over 4s T},e^{-{1\over4sT^2}}\right),
\end{equation}
where $\vartheta_3(u,q)$ is the elliptictheta function,
\begin{equation}
\vartheta_3\left(u,q\right)\equiv 1+2\sum_{n=1}^\infty q^{n^2}\cos(2nu).
\end{equation}
The general behavior of ${\cal P}_{\rm c}(\tau)$ at finite $T$ is shown in Fig. \ref{Point_Charged_Thermal_Correlation}. 
%---------------------------------------------------------------------
%---------------------------------------------------------------------
At finite $T$, ${\cal P}_{\rm c}(\tau)$ is no longer a pure decreasing function, 
but an oscillating function with periodicity $\beta=1/T$. Therefore, the zero-temperature decaying behavior ${\cal P}_{\rm c}(\tau)\sim e^{-\tau m_{\pi^\pm}}$ cannot be used to extract the charged pion mass. The maxima decay fast due to the term $\exp{[-\tau^2/(4s)]}$. They are located exactly at $\tau=n\beta$ ($n=0,1,2,...$). Therefore, at large $n$, we have
\begin{equation}
{\cal P}_{\rm c}(\tau=n\beta)\sim e^{-n\beta m_{\pi^\pm}}.
\end{equation}
This method should maintain gauge independence and gives $m_{\pi^\pm}({\rm e}B)=\sqrt{m_\pi^2+{\rm e}B}$. However, since $n$ is not a continuous variable, it is not quite useful for a numerical extraction of the mass. On the other hand, the minima go smoothly and may be used to extract the mass numerically. Near the first minimum $\tau=\beta/2$, the correlation function behaves as
\begin{equation}\label{Mc-cosh}
{\cal P}_{\rm c}(\tau)\sim \cosh\left[m_{\pi^\pm}(\tau-\beta/2)\right].
\end{equation}
This behavior was used to extract the hadron masses in the lattice QCD calculations~\cite{Bali:2017ian,Ding:2020jui}. Here we also use this function to extract the charged pion mass at finite temperature
from Eq. (\ref{point_particle_model_T}). The numerical results extracted from the data around the first minimum are shown in Fig.~\ref{Thermal}.
%---------------------------------------------------------------------
%---------------------------------------------------------------------
In the free point particle model without interaction, the charged pion mass is always $\sqrt{m_\pi^2+{\rm e}B}$ at arbitrary temperature. Therefore, the gauge dependence of the charged pion mass becomes severe at high temperature. The breaking of gauge independence is due to the mixing of higher Landau levels with the LLL at finite temperature. However, at sufficiently low temperature, gauge independence still remains.  At larger ${\rm e}B$, the higher Landau levels are suppressed compared to the LLL, and hence the temperature window where the gauge independence is nearly guaranteed is extended to higher temperature. This can be understood by taking ${\rm e}B\rightarrow \infty$ in Eq.~\eqref{point_particle_model_T}. In this case, $\coth({\rm e}Bs)\simeq 1$ and hence the gauge-dependent term $\sqrt{\xi^2+\coth({\rm e}Bs)}\simeq\sqrt{\xi^2+1}$. Therefore, the extraction scheme of Eq. (\ref{Mc-cosh}) breaks down when $T$ is not much smaller than 
$\sqrt{m_\pi^2+{\rm e}B}$. In this case, it may be better to introduce a small isospin chemical potential and extract the charged pion mass from the isospin density as it is more sensitive to the energy.

%%%%%%%%%%%%%%%%%%%%%%%%%%%%%%%%%%%%%%%%%%%%%%%%%%%%%%%%%%%
\subsubsection{Charged Pions as Composite Particles in the NJL Model}
%%%%%%%%%%%%%%%%%%%%%%%%%%%%%%%%%%%%%%%%%%%%%%%%%%%%%%%%%%%
	
\begin{figure}[H]
	\centering
	\includegraphics[width=0.4\textwidth]{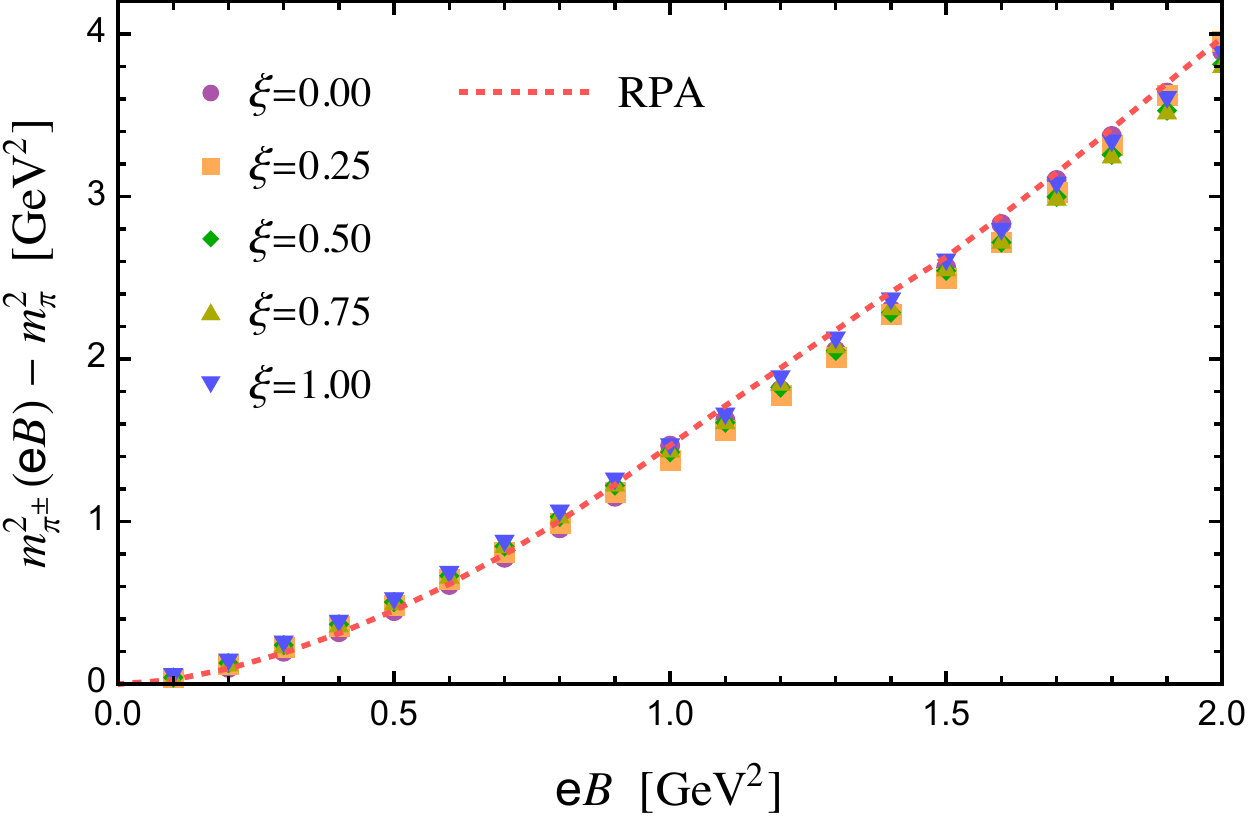}
	\caption{The charged pion mass squared as a function of ${\rm e}B$  in the NJL model, with the gauge constant $\xi$ varying from $0$ to $1$. The result from the momentum-space RPA (red dotted line), which discards the Schwinger phase, is also shown for comparison.}
\label{charged_pion}
\end{figure}

In the NJL model, the charged pions are composite bosons. If we choose the vector potential ${\bf A}={\bf A}_\perp$, we can extract their mass at zero temperature from Eq. (\ref{decay_charged_pion}) at large $\tau$. The results for different gauge parameters $\xi$ are illustrated in Fig.~\ref{charged_pion}.   Regardless of the numerical accuracy, the gauge independence seems to be still well satisfied. Our numerical results also agree well with the result from the momentum-space RPA. 
%---------------------------------------------------------------------
%---------------------------------------------------------------------

\begin{figure}[H]
	\centering
\includegraphics[width=0.4\textwidth]{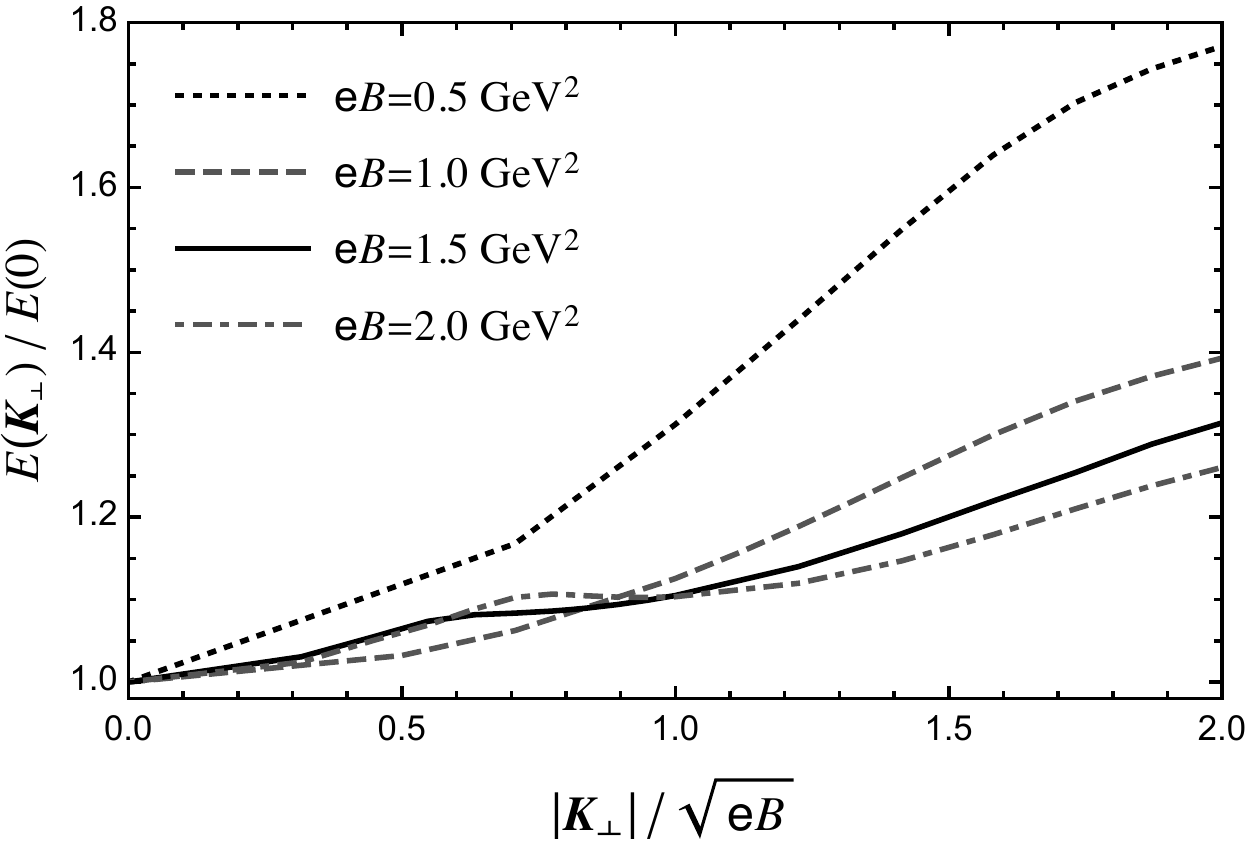}
\caption{The dispersion $E({\bf K}_\perp)$ of the charged pions for different values of ${\rm e}B$ in the NJL model.}
\label{K_dependence}
\end{figure}

The gauge independence can be exactly proven as follows. For the present choice of the vector potential ${\bf A}={\bf A}_\perp$, we have
\begin{eqnarray}
{\cal P}_{\rm c}(\tau)
=\int_{0}^{\infty}\frac{\mathrm{d}\mathbf{K}_\perp^2}{{\rm e}B\xi}J_0\left(\frac{\mathbf{K}_\perp^2}{{\rm e}B\xi}\right)\left[\int_{-\infty}^{\infty}\frac{\mathrm{d}K_4}{2\pi}e^{-iK_4\tau}\bar{\cal D}_{\rm c}(K_4,\mathbf{K}_\perp,K_3=0)\right]. 
\end{eqnarray}
The integral over $K_4$ can be converted to a contour as we have done for the neutral pion. For $\tau\rightarrow\infty$, it is going to pick up the lowest branch of the poles for $K_0$ determined by the pole equation
\begin{eqnarray}
1+2g\bar{\Pi}_{\rm c}(K_4\rightarrow -iK_0,\mathbf{K}_\perp,K_3=0)=0.
\end{eqnarray}
For convenience, we denote this ${\bf K}_\perp$-dependent pole as $E({\bf K}_\perp)$. Therefore, for $\tau\rightarrow\infty$, the correlation function ${\cal P}_{\rm c}(\tau)$ goes as
\begin{eqnarray}
{\cal P}_{\rm c}(\tau)
&\sim& \int_{0}^{\infty}\frac{\mathrm{d}\mathbf{K}_\perp^2}{{\rm e}B\xi}J_0\left(\frac{\mathbf{K}_\perp^2}{{\rm e}B\xi}\right)\exp{\left[-\tau E({\bf K}_\perp)\right]}\nonumber\\ 
&=&\frac{1}{{\rm e}B\xi\pi}\int_{-\infty}^{\infty}\mathrm{d}K_1\int_{-\infty}^{\infty}\mathrm{d}K_2J_0\left(\frac{K_1^2+K_2^2}{{\rm e}B\xi}\right)\exp{\left[-\tau E(K_1,K_2)\right]}
\end{eqnarray}
While the integral over ${\bf K}_\perp=(K_1,K_2)$ cannot be carried out analytically, we can use the Laplace method for $\tau\rightarrow\infty$. Supposing the global minimum of the dispersion $E(K_1,K_2)$ on the $K_1-K_2$ plane is located at $(K_1,K_2)=(u_1,u_2)$, i.e.,
\begin{eqnarray}
\min\{E(K_1,K_2)\}=E(u_1,u_2).
\end{eqnarray}
Using the Laplace method, we arrive at the large-$\tau$ behavior
\begin{eqnarray}
{\cal P}_{\rm c}(\tau)\sim \exp{\left[-\tau E(u_1,u_2)\right]},
\end{eqnarray}
where the prefactors are irrelevant for the mass and are not shown here. Therefore, the mass is exactly determined by the minimum of the dispersion $E({\bf K}_\perp)$ solved from the pole equation. 
If the minimum is located at ${\bf K}_\perp=0$, this justifies the momentum-space RPA which discards the Schwinger phase and solves the pole mass at ${\bf K}=0$. In Fig.~\ref{K_dependence}, the numerical results show that all the minimum is always located at ${\bf K}_\perp=0$ for different values of ${\rm e}B$.
	
%---------------------------------------------------------------------
%---------------------------------------------------------------------

For a general choice of the vector potential ${\bf A}$, we can also show that mass is exactly determined by the minimum of the dispersion $E({\bf K})$, solved from the pole equation in the momentumm space RPA. In this case, we can directly start from the correlation function
\begin{eqnarray}
{\cal D}_{\rm c}(\tau,{\bf r};\tau^\prime,{\bf r}^\prime)
&=& e^{i\Phi\left(\mathbf{r},\mathbf{r}^\prime\right)}\int\frac{\mathrm{d}^4K}{(2\pi)^4}e^{-iK_4(\tau-\tau^\prime)}e^{i\mathbf{K}\cdot\left(\mathbf{r}-{\bf r}^\prime\right)}\bar{\cal D}_{\rm c}(K)\nonumber\\
&=& e^{i\Phi\left(\mathbf{r},\mathbf{r}^\prime\right)}\int\frac{\mathrm{d}^3{\bf K}}{(2\pi)^3}e^{i\mathbf{K}\cdot\left(\mathbf{r}-{\bf r}^\prime\right)}
\int_{-\infty}^\infty\frac{{\rm d}K_4}{2\pi}e^{-iK_4(\tau-\tau^\prime)}\bar{\cal D}_{\rm c}(K_4,{\bf K}).
\end{eqnarray}
Due to the translational invariance in the temporal direction, we can set $\tau^\prime=0$. For $\tau\rightarrow\infty$, the integral over $K_4$ picks up the lowest branch of the poles, $K_0=E({\bf K})$, determined by the pole equation in the momentum space RPA,
\begin{eqnarray}\label{MRPA-K}
1+2g\bar{\Pi}_{\rm c}(K_4\rightarrow -iK_0,\mathbf{K})=0.
\end{eqnarray}
Therefore, for $\tau\rightarrow\infty$, the correlation function goes as
\begin{eqnarray}
{\cal D}_{\rm c}(\tau,{\bf r},{\bf r}^\prime)\sim e^{i\Phi\left(\mathbf{r},\mathbf{r}^\prime\right)}\int\frac{\mathrm{d}^3{\bf K}}{(2\pi)^3}e^{i\mathbf{K}\cdot\left(\mathbf{r}-{\bf r}^\prime\right)}
e^{-\tau E({\bf K})}
\end{eqnarray}
Supposing the global minimum of the dispersion $E({\bf K})$ is located at ${\bf K}={\bf u}$, i.e.,
\begin{eqnarray}
\min\{E({\bf K})\}=E({\bf u}).
\end{eqnarray}
Using the Laplace method, we arrive at the large-$\tau$ behavior
\begin{eqnarray}
{\cal D}_{\rm c}(\tau,{\bf r},{\bf r}^\prime)\sim e^{i\Phi\left(\mathbf{r},\mathbf{r}^\prime\right)}e^{i\mathbf{u}\cdot\left(\mathbf{r}-{\bf r}^\prime\right)}e^{-\tau E({\bf u})},
\end{eqnarray}
where some prefactors are not shown. Here we see clearly that the Schwinger phase can be discarded if we only need to determine the mass. Normally, the minimum of the dispersion is located at 
${\bf K}=0$, i.e., ${\bf u}=0$. This justifies the validity of the momentum-space RPA which solves the pole mass from Eq. (\ref{MRPA-K}) at ${\bf K}=0$.  This argument can be extended to other 
charged mesons in a constant magnetic field, such as charged rho mesons~\cite{Cao:2019res}.

%%%%%%%%%%%%%%%%%%%%%%%%%%%%%%%%%%%%%%%%%%%%%%%%%%%%%%%%%%%%%%%%%%%%%%%%%%%%%%%%%%%%%%%%%%%%%%%%%%%%%
\section{Summary}\label{sec5}
%%%%%%%%%%%%%%%%%%%%%%%%%%%%%%%%%%%%%%%%%%%%%%%%%%%%%%%%%%%%%%%%%%%%%%%%%%%%%%%%%%%%%%%%%%%%%%%%%%%%%	
We have investigated the masses of neutral and charged pions in a constant magnetic field within the NJL model. To fully take into account the Schwinger phase, we start from the meson correlation functions in the coordinate space and determine the meson masses from the exponential behavior at large imaginary time. Within this imaginary-time method, we show numerically and analytically that the mass of the charged pions is independent of the choice of the vector potential for the constant background magnetic field. We also demonstrate that the previously used momentum-space RPA~\cite{Liu:2018zag,Cao:2019res}, which simply discarded the Schwinger phases and determined the meson mass as the pole at zero momentum, is actually equivalent to the imaginary-time method used in this work.

\begin{acknowledgments}
J. L. and L. H. are supported by the National Natural Science Foundation of China under Grant Nos. 11775123 and 11890712. G.C. is supported by the National Natural Science Foundation of China with Grant No. 11805290 and Young Teachers Training Program of Sun Yat-sen University with Grant No. 19lgpy282.
\end{acknowledgments}

\appendix
\section{Illustration of Eq.~(\ref{redefinitions})}
\label{Appendix}
For any two given space-time points $Z$ and $Z^\prime$, the correlation function $\mathcal{D}_\text{c}\left(Z,Z^\prime\right)$ is given by Eq.~(\ref{Dc-definition}). Using the expression (\ref{Wmf}), we have
\begin{equation}
	\begin{split}
		\mathcal{D}_\text{c}\left(Z,Z^\prime\right)&=\frac{\delta^2\mathcal{W}^{(2)}_\text{MF}\left[J\right]}{\delta J_+\left(Z\right)\delta J_-\left(Z^\prime\right)} \\
		&=-\frac{\delta}{\delta J_+\left(Z\right)}\int\text{d}^4X\int\text{d}^4X'\begin{pmatrix} J_+\left(X\right) &V_+\left(X\right)  \end{pmatrix}
		\begin{pmatrix} \Pi\left(X,X'\right) & -\Pi\left(X,X'\right) \\ -\Pi\left(X,X'\right) & \frac{\delta^{(4)}\left(X-X'\right) }{2g}+\Pi\left(X,X'\right) \end{pmatrix}\begin{pmatrix} 1 \\  \frac{\delta V_-\left(X'\right)}{\delta J_-\left(X^\prime\right)}\end{pmatrix}\delta^{(4)}\left(X^\prime-Z^\prime\right)\\
		&=-\frac{\delta}{\delta J_+\left(Z\right)}\int\text{d}^4X\int\text{d}^4X'\begin{pmatrix}  J_+\left(X\right) &V_+\left(X\right)  \end{pmatrix}e^{i\Phi\left(X,X'\right)}
		\begin{pmatrix} \bar\Pi\left(X-X'\right) & -\bar\Pi\left(X-X'\right) \\ -\bar\Pi\left(X-X'\right) & \frac{\delta^{(4)}\left(X-X'\right) }{2g}+\bar\Pi\left(X-X'\right) \end{pmatrix}\begin{pmatrix} 1 \\  \frac{\delta V_-\left(X'\right)}{\delta  J_-\left(X'\right)}\end{pmatrix}\delta^{(4)}\left(X^\prime-Z'\right)\\
		&=-\frac{\delta}{\delta J_+\left(Z\right)}\int\text{d}^4X\int\text{d}^4X'\begin{pmatrix}  J_+\left(X\right) &V_+\left(X\right)  \end{pmatrix}e^{i\Phi\left(X,Z'\right)}
		\begin{pmatrix} \bar\Pi\left(X-X'\right) & -\bar\Pi\left(X-X'\right) \\ -\bar\Pi\left(X-X'\right) & \frac{\delta^{(4)}\left(X-X'\right) }{2g}+\bar\Pi\left(X-X'\right) \end{pmatrix}\begin{pmatrix} 1 \\  \frac{\delta V_-\left(X'\right)}{\delta  J_-\left(X'\right)}\end{pmatrix}\delta^{(4)}\left(X^\prime-Z'\right)\\
		&=-\frac{\delta}{\delta J_+\left(Z\right)}\int\text{d}^4X\int\text{d}^4X'\begin{pmatrix} \bar J_+\left(X\right) &\bar V_+\left(X\right)  \end{pmatrix}
		\begin{pmatrix} \bar\Pi\left(X-X'\right) & -\bar\Pi\left(X-X'\right) \\ -\bar\Pi\left(X-X'\right) & \frac{\delta^{(4)}\left(X-X'\right) }{2g}+\bar\Pi\left(X-X'\right) \end{pmatrix}\begin{pmatrix} 1 \\  \frac{\delta \bar V_-\left(X'\right)}{\delta \bar J_-\left(X'\right)}\end{pmatrix}\delta^{(4)}\left(X^\prime-Z'\right)\\
		&=-\frac{\delta}{\delta J_+\left(Z\right)\delta \bar J_-\left(Z'\right)}\int\text{d}^4X\int\text{d}^4X'\begin{pmatrix} \bar J_+\left(X\right) &\bar V_+\left(X\right)  \end{pmatrix}
		\begin{pmatrix} \bar\Pi\left(X-X'\right) & -\bar\Pi\left(X-X'\right) \\ -\bar\Pi\left(X-X'\right) & \frac{\delta^{(4)}\left(X-X'\right) }{2g}+\bar\Pi\left(X-X'\right) \end{pmatrix}\begin{pmatrix} \bar J_-\left(X'\right) \\   \bar V_-\left(X'\right)\end{pmatrix}\\
		&=\frac{\delta^2\overline{\mathcal{W}}^{(2)}_\text{MF}\left[\bar J\right]}{\delta J_+\left(Z\right)\delta \bar J_-\left(Z'\right)}=e^{i\Phi\left(Z,Z'\right)}\frac{\delta^2\overline{\mathcal{W}}^{(2)}_\text{MF}\left[\bar J\right]}{\delta \bar J_+\left(Z\right)\delta \bar J_-\left(Z'\right)}.
	\end{split}
\end{equation}
Notice that the generating functional is changed in the last line. The phase term in the original generating functional $\mathcal{W}_\text{MF}^{(2)}\left[J\right]$ is the Schwinger phase $\exp\left[i\Phi\left(X,X'\right)\right]$, while it becomes $\exp\left[i\Phi\left(X,Z'\right)+i\Phi\left(Z',X'\right)\right]$ in the modified version $\overline{\mathcal{W}}_\text{MF}^{(2)}\left[\bar J\right]$. 
Fortunately, the new phase term yields to the Schwinger phase when performing the variational derivative with respect to the source $\bar{J}_-$ located at $Z'$.  The last line provides an 
equivalent functional definition of the charged pion correlation function, which is convenient for us to perform the calculation.

%----------------------------------------------------------------------------------
%\bibliographystyle{apsrev}
%\bibliographystyle{apsrev4-2}
%\bibliography{Reference}
	
\end{document}